\documentclass[sigconf,nonacm]{acmart} 
\AtBeginDocument{%
  }

\acmConference[Preprint]{Under review at the 34th ACM International Conference on Multimedia (MM '26)}{November 10--14, 2026}{Rio de Janeiro, Brazil}

\usepackage{graphicx}     
\usepackage{multirow}
\usepackage[table]{xcolor}
\usepackage{booktabs}
\usepackage{threeparttable}
\usepackage{makecell}
\usepackage{siunitx}
\usepackage{algorithm}
\usepackage{algpseudocode}
\usepackage{enumitem}
\usepackage{bm}
\usepackage{pifont} 
\usepackage{colortbl,xcolor}
\definecolor{lightgray}{gray}{0.9} 

\begin{document}
\title{\textit{SatFusion}: A Unified Framework for Enhancing Remote Sensing Images via Multi-Frame and Multi-Source Images Fusion}

\author{Yufei Tong}
\affiliation{%
  \institution{Zhejiang University}
  \city{Hangzhou}
  \country{China}}
\email{yufeitong@zju.edu.cn}

\author{Guanjie Cheng\textsuperscript{\textdagger}}
\affiliation{%
  \institution{Zhejiang University}
  \city{Hangzhou}
  \country{China}}
\email{chengguanjie@zju.edu.cn}
\thanks{\textsuperscript{\textdagger} Guanjie Cheng is the corresponding author.}

\author{Peihan Wu}
\affiliation{%
  \institution{Zhejiang University}
  \city{Hangzhou}
  \country{China}}
\email{22321313@zju.edu.cn}

\author{Feiyi Chen}
\affiliation{%
  \institution{Zhejiang University}
  \city{Hangzhou}
  \country{China}}
\email{chenfeiyi@zju.edu.cn}

\author{Xinkui Zhao}
\affiliation{%
  \institution{Zhejiang University}
  \city{Hangzhou}
  \country{China}}
\email{zhaoxinkui@zju.edu.cn}

\author{Shuiguang Deng}
\affiliation{%
  \institution{Zhejiang University}
  \city{Hangzhou}
  \country{China}}
\email{dengsg@zju.edu.cn}
  
\renewcommand{\shortauthors}{Tong et al.}

\begin{abstract}
High-quality remote sensing (RS) image acquisition is fundamentally constrained by physical limitations. While Multi-Frame Super-Resolution (MFSR) and Pansharpening address this by exploiting complementary information, they are typically studied in isolation: MFSR lacks high-resolution (HR) structural priors for fine-grained texture recovery, whereas Pansharpening relies on upsampled low-resolution (LR) inputs and is sensitive to noise and misalignment.
In this paper, we propose \textbf{\textit{SatFusion}}, a novel and unified framework that seamlessly bridges multi-frame and multi-source RS image fusion. \textbf{\textit{SatFusion}} extracts HR semantic features by aggregating complementary information from multiple LR multispectral frames via a Multi-Frame Image Fusion (MFIF) module, and integrates fine-grained structural details from an HR panchromatic image through a Multi-Source Image Fusion (MSIF) module with implicit pixel-level alignment. 
To further alleviate the lack of structural priors during multi-frame fusion, we introduce an advanced variant, \textbf{\textit{SatFusion*}}, which integrates a panchromatic-guided mechanism into the MFIF stage. Through structure-aware feature embedding and transformer-based adaptive aggregation, \textbf{\textit{SatFusion*}} enables spatially adaptive feature selection, strengthening the coupling between multi-frame and multi-source representations. 
Extensive experiments on four benchmark datasets validate our core insight: synergistically coupling multi-frame and multi-source priors effectively resolves the fragility of existing paradigms, delivering superior reconstruction fidelity, robustness, and generalizability.

\end{abstract}


  
\maketitle

\section{Introduction}
\label{sec:1}
High-quality remote sensing (RS) imagery is crucial for diverse downstream applications~\cite{do2024c,li2020review,neyns2022mapping,wellmann2020remote,xu2023ai,yang2013role,yuan2020deep}, yet its acquisition is fundamentally constrained by sensor hardware limits~\cite{pohl1998review, loncan2015hyperspectral}. To overcome these constraints, image fusion has evolved along two primary trajectories: Multi-Frame Super-Resolution (MFSR)~\cite{bhat2021deep,deudon2020highres,wei2023towards,salvetti2020multi,an2022tr,di2025qmambabsr}, which aggregates complementary information from multiple low-resolution (LR) frames, and Pansharpening~\cite{deng2022machine,loncan2015hyperspectral,meng2020large,thomas2008synthesis,he2023multiscale, vivone2020new,zhu2023probability,huang2025general}, which fuses a high-resolution (HR) panchromatic (PAN) image with an LR multispectral (MS) image.

\begin{figure}
    \centering
    \includegraphics[width=1.0\linewidth]{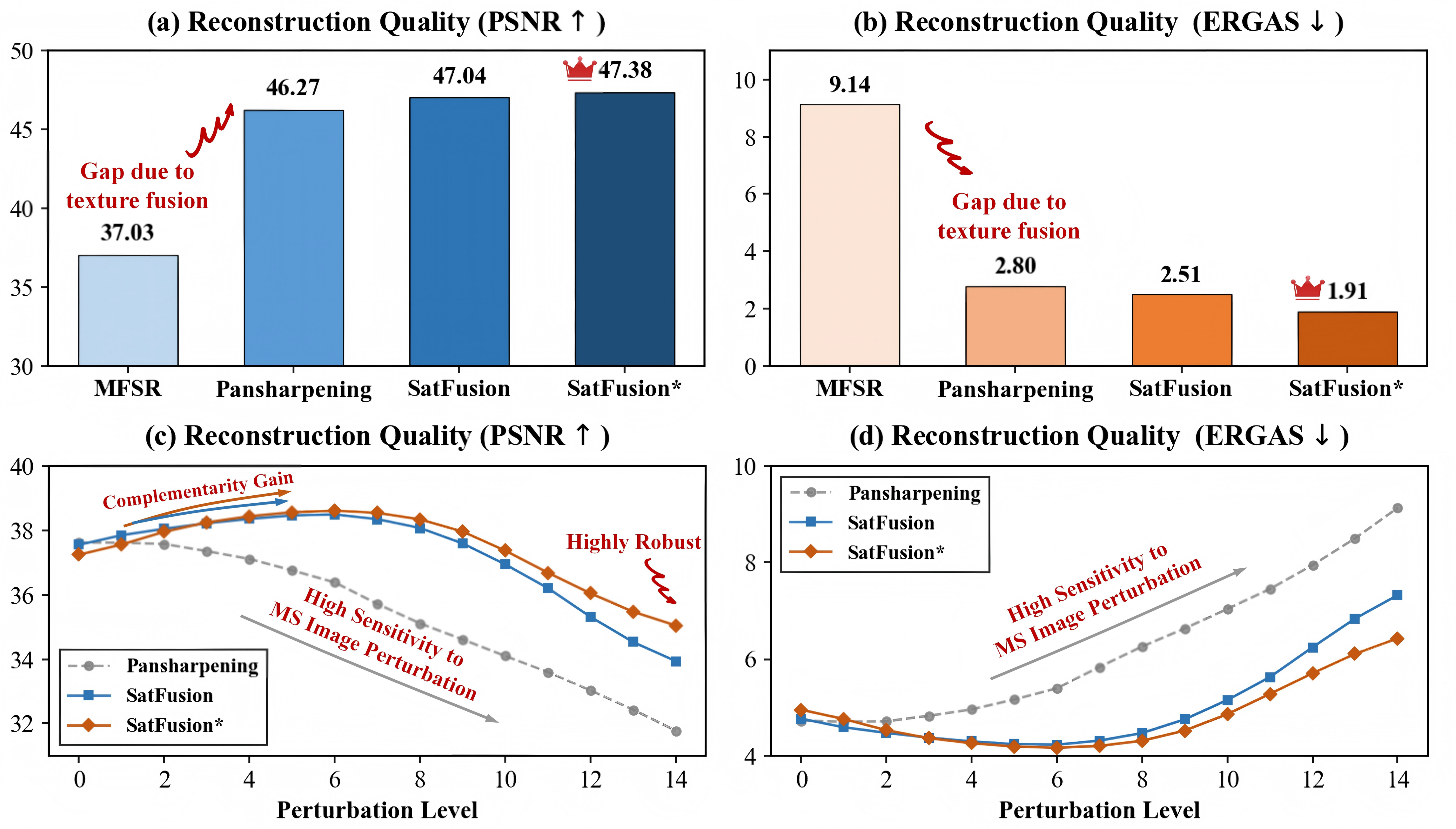}
    \caption{
        \textbf{Motivation and superiority of our proposed framework.} 
        \textbf{(a-b)} Visual comparison on WorldStrat dataset. The performance gap between MFSR and Pansharpening emphasizes the critical need for PAN structural priors. 
        \textbf{(c-d)} Robustness against increasing compound perturbations (blur, noise, and misalignment) on the QB dataset. Unlike traditional Pansharpening which suffers severe degradation, our \textit{SatFusion} and \textit{SatFusion*} maintain superior robustness by effectively leveraging multi-frame complementary information.
    }
    \label{fig:1}
\end{figure}

Despite their respective successes, these two paradigms are typically studied in isolation, leaving fundamental challenges unresolved.
First, \textbf{MFSR lacks HR structural priors}. While multi-frame inputs provide complementary sub-pixel information, the absence of high-frequency guidance fundamentally limits the recovery of fine-grained textures, resulting in a persistent performance bottleneck (as evidenced by the gap in Fig.~\ref{fig:1}(a-b)).
Second, \textbf{Pansharpening is notoriously sensitive to noise and spatial misalignment}. Pansharpening requires explicitly upsampling the LR MS image to match the PAN resolution prior to fusion~\cite{masi2016Pansharpening,yang2017pannet,he2019Pansharpening,deng2020detail,xing2024empower,zhong2024ssdiff,wang2025adaptive,do2025pan,wang2025mmmamba}. This explicit upsampling inevitably introduces interpolation artifacts and magnifies noise. Consequently, under real-world perturbations (e.g., imaging noise or inter-source shifts), traditional Pansharpening suffers from severe blurring and performance collapse, as illustrated in Fig.~\ref{fig:1}(c-d). Neither paradigm alone can robustly process the massive, low-quality, yet complementary data generated by modern satellite constellations~\cite{kim2021satellite,kothari2020final,lofqvist2020accelerating,wang2023satellite}.

To break this isolation, we propose \textbf{\textit{SatFusion}}, the first unified framework designed to seamlessly integrate multi-frame and multi-source RS image fusion. Instead of relying on fragile explicit upsampling, \textit{SatFusion} employs a Multi-Frame Image Fusion (MFIF) module to extract semantic features from multiple LR MS frames, while a Multi-Source Image Fusion (MSIF) module concurrently injects fine-grained structural priors from the HR PAN image. This synergistic design simultaneously addresses both fundamental bottlenecks: it provides the crucial \textbf{HR structural guidance} lacking in traditional MFSR for fine-grained texture recovery, and achieves \textbf{implicit pixel-level alignment} to circumvent the artifact amplification inherent in Pansharpening. Importantly, \textit{SatFusion} provides a standardized and extensible feature interface, allowing existing MFSR and Pansharpening methods to be naturally embedded.

Furthermore, recognizing that practical multi-frame MS inputs are often plagued by spatial misalignments and variable sequence lengths, we propose an advanced formulation, \textbf{\textit{SatFusion*}}, which introduces a PAN-guided mechanism into the MFIF module. By leveraging structure-aware feature embedding and transformer-based adaptive aggregation, the stable geometric structure of the PAN image serves as a reliable spatial reference. This mechanism guides the spatially adaptive selection of multi-frame features, forcing the structural constraints of the PAN image to directly inform multi-frame aggregation decisions. Simultaneously, the Transformer architecture inherently supports arbitrary sequence lengths. As shown in Fig.~\ref{fig:1}, \textit{SatFusion*} significantly enhances both reconstruction quality and robustness against input perturbations.

The main contributions of this work are summarized as follows:
\begin{itemize}[leftmargin=1.5em]
\item We reveal the inherent structural complementarity between MFSR and Pansharpening and propose \textbf{\textit{SatFusion}}, the first unified framework for enhancing RS image via multi-frame and multi-source image fusion. To the best of our knowledge, this is the first work to investigate the joint optimization of multi-frame and multi-source RS images within a unified framework.
\item We introduce an advanced variant, \textbf{\textit{SatFusion*}}. By incorporating structural priors into the MFIF module, \textit{SatFusion*} enables spatially adaptive multi-frame feature aggregation, strengthening the coupling between multi-frame and multi-source features and improving model generalization across diverse input scenarios.
\item Extensive experiments on four datasets validate our core insight: unifying multi-frame and multi-source information fundamentally shatters the limitations of isolated paradigms. This unified framework not only yields superior reconstruction fidelity but also empowers \textit{SatFusion*} with exceptional generalizability—effectively mitigating noise perturbations while natively adapting to arbitrary inference frame counts—offering a new solution for practical RS scenarios.
\end{itemize}

\section{Related Work}
\label{sec:2}

\subsection{Multi-Frame Super-Resolution (MFSR)}
\label{sec:2.1}
Unlike Single-Image Super-Resolution (SISR) which relies purely on learned image priors, MFSR~\cite{bhat2021deep,deudon2020highres,molini2019deepsum,salvetti2020multi,an2022tr,di2025qmambabsr} reconstructs HR images by exploiting complementary sub-pixel information across multiple LR observations (Fig.~\ref{fig:2}(a)). In natural burst photography, methods typically employ optical flow or attention mechanisms to aggregate slightly misaligned frames~\cite{bhat2021deep, wei2023towards, di2025qmambabsr}. In RS scenarios, where MFSR is commonly referred to as Multi-Image Super-Resolution (MISR), the challenge is exacerbated by longer temporal intervals and complex orbital variations. To address this, previous works have explored various feature fusion strategies, ranging from 2D/3D convolutions (e.g., HighRes-Net~\cite{deudon2020highres, razzak2023multi}, DeepSUM~\cite{molini2019deepsum}, RAMS~\cite{salvetti2020multi}) to transformer-based spatial-temporal attention architectures (e.g., TR-MISR~\cite{an2022tr}). 
\textbf{Limitation:} Despite these structural advances, existing MFSR methods operate exclusively on LR inputs. Without explicit HR structural guidance, their ability to reconstruct fine-grained, high-frequency spatial textures remains fundamentally bottlenecked.

\subsection{Pansharpening}
\label{sec:2.2}
\begin{figure}
    \centering
    \includegraphics[width=\linewidth]{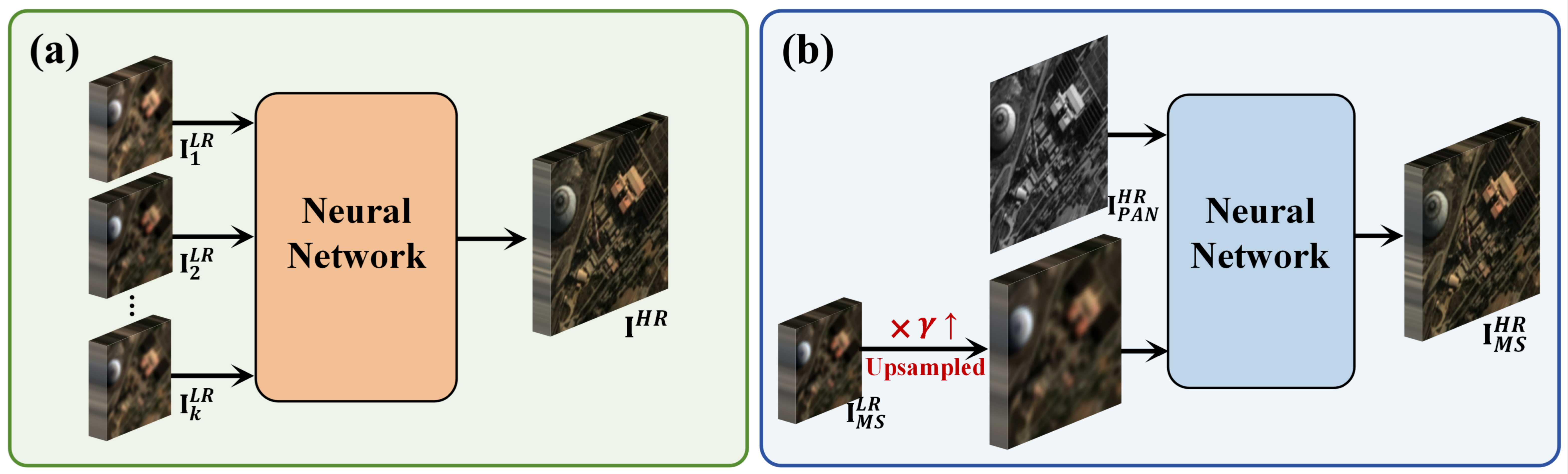}
    \caption{
        The typical paradigms of (a) Multi-Frame Super-Resolution (MFSR) and (b) Pansharpening. Our work breaks this isolated design by unifying both paradigms.
    }
    \label{fig:2}
\end{figure}

Pansharpening focuses on the spatial enhancement of an LR MS image guided by an HR PAN image acquired over the same scene (Fig.~\ref{fig:2}(b)). Driven by deep learning, Pansharpening has evolved from early CNN architectures (e.g., PNN~\cite{masi2016Pansharpening}, PanNet~\cite{yang2017pannet}, FusionNet~\cite{deng2020detail}) to more complex paradigms~\cite{he2019Pansharpening, peng2023u2net}. Recently, diffusion models~\cite{kim2025u,meng2023pandiff,xing2024empower,zhong2024ssdiff,xing2025dual} have been introduced for iterative detail injection, while state-space models and adaptive convolutions~\cite{jin2022lagconv,duan2024content} (e.g., Pan-Mamba~\cite{he2025pan}, ARConv~\cite{wang2025adaptive}) have been explored to capture long-range dependencies and anisotropic structural patterns.
\textbf{Limitation:} A critical flaw in current Pansharpening pipelines is their reliance on explicitly upsampling the LR MS image to match the PAN resolution prior to fusion~\cite{masi2016Pansharpening,yang2017pannet,he2019Pansharpening,deng2020detail,xing2024empower,zhong2024ssdiff,wang2025adaptive,do2025pan,wang2025mmmamba}. This pre-processing step severely amplifies sensor noise and exacerbates inter-modal misalignment, causing significant performance degradation under real-world perturbations.

\begin{figure*}[t]
    \centering
    \includegraphics[width=0.95\linewidth]{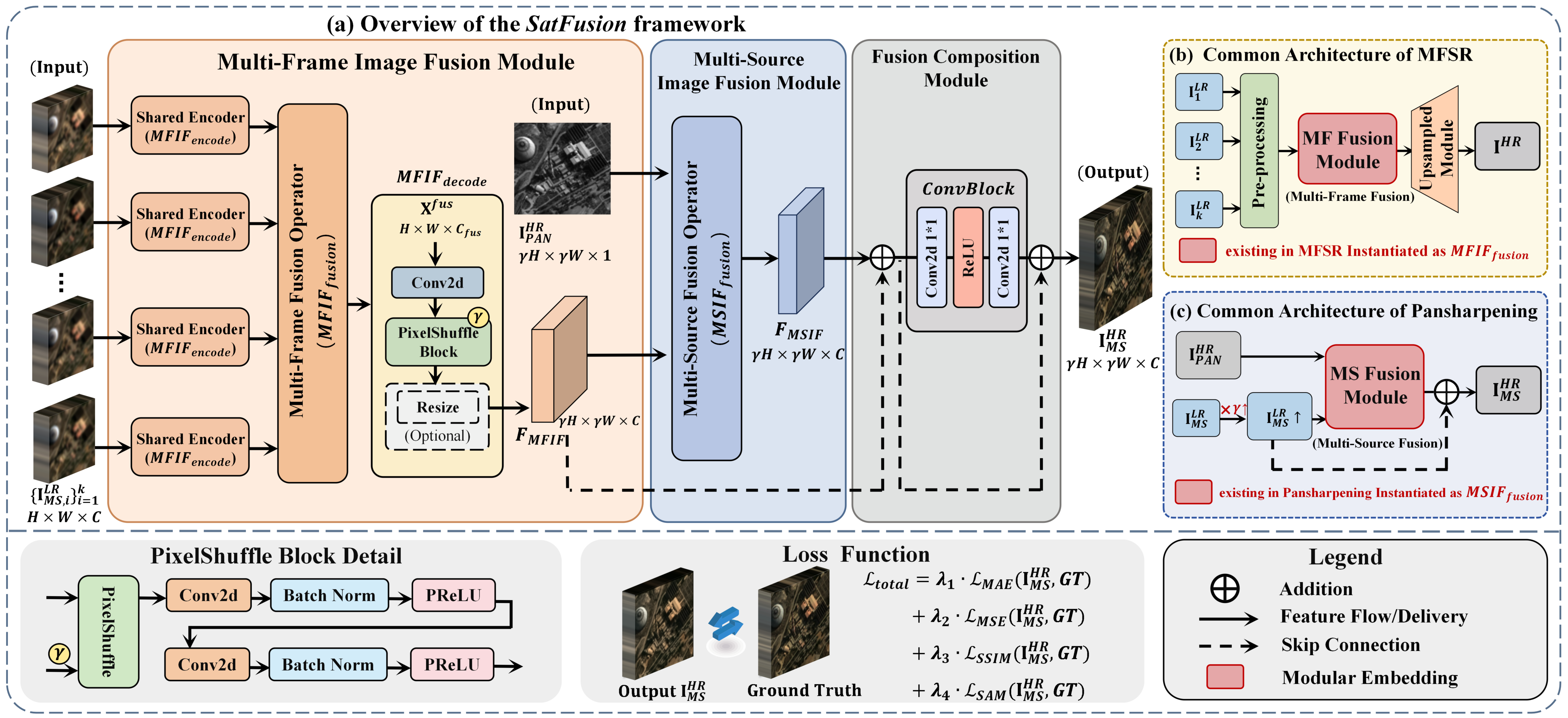}
    \caption{
        (a) SatFusion framework overview. The MFIF module aggregates multi-frame LR MS inputs ($\{\mathbf{I}_{MS,i}^{LR}\}$) into HR deep semantic features, and the MSIF module injects fine-grained textures from HR PAN ($\mathbf{I}_{PAN}^{HR}$). The Fusion Composition module merges features for final reconstruction, guided by joint loss functions. (b-c) Our unified framework enables existing MFSR and Pansharpening components to be naturally embedded.
    }
    \label{fig:3}
\end{figure*}

\subsection{Motivations}
\label{sec:2.3}
As analyzed above, MFSR and Pansharpening possess highly complementary strengths and weaknesses. MFSR leverages multi-frame redundancy, making it inherently robust to single-frame noise, yet it suffers from blurry reconstructions due to the lack of HR priors. Conversely, Pansharpening provides sharp spatial structures but is highly fragile to noise and misalignment. 
Surprisingly, the joint optimization of these two tasks remains largely unexplored. The core motivation of our work is to bridge this gap. By proposing \textit{SatFusion}, we eliminate the fragile explicit upsampling in Pansharpening by using multi-frame MS features for implicit alignment, while simultaneously breaking the MFSR performance ceiling by injecting PAN structural priors. Furthermore, our advanced \textit{SatFusion*} introduces a PAN-guided mechanism directly into the multi-frame aggregation stage, ensuring spatially adaptive fusion that is highly robust to varying frame counts and severe input degradation.

\section{Methodology}
\label{sec:3}
In this section, we formulate the joint multi-frame and multi-source RS image fusion task (Section~\ref{sec:3.1}) and detail the proposed \textit{SatFusion} framework (Section~\ref{sec:3.2}), its advanced variant \textit{SatFusion*} (Section~\ref{sec:3.3}), and the optimization objectives (Section~\ref{sec:3.4}). Detailed network architectures and dimensional transformations for all modules are provided in Appendix~\ref{app:architecture}.

\subsection{Problem Formulation}
\label{sec:3.1}
To better align with real-world satellite imaging scenarios, we formulate a unified task: reconstructing a high-quality, HR MS image by jointly fusing multiple LR MS frames and a single HR PAN image of the same scene. Given $k$ LR MS images $\{\mathbf{I}_{MS, i}^{LR}\}_{i=1}^{k} \in \mathbb{R}^{k \times H \times W \times C}$ and an HR PAN image $\mathbf{I}_{PAN}^{HR} \in \mathbb{R}^{\gamma H \times \gamma W \times 1}$, our goal is to learn a mapping function $\mathcal{F}(\cdot)$ to reconstruct the HR MS image $\mathbf{I}_{MS}^{HR} \in \mathbb{R}^{\gamma H \times \gamma W \times C}$:
\begin{equation}
\mathbf{I}_{MS}^{HR} = \mathcal{F}\Big(\{\mathbf{I}_{MS, i}^{LR}\}_{i=1}^{k}, \mathbf{I}_{PAN}^{HR}\Big),
\end{equation}
where $\gamma$ denotes the spatial upscaling factor, and $C$ represents the number of MS spectral channels.

\subsection{\textit{SatFusion}: A Unified Framework}
\label{sec:3.2}
The primary goal of \textit{SatFusion} is to break the isolated design para\-digms of MFSR and Pansharpening. By doing so, it provides a highly extensible blueprint that circumvents the fragile explicit MS upsampling required in traditional Pansharpening. As illustrated in Fig.~\ref{fig:3}(a), it consists of three collaborative modules.

\textbf{1) Multi-Frame Image Fusion (MFIF):} 
Unlike conventional Pansharpening pipelines that naively upsample the raw LR MS image---inevitably magnifying sensor noise---our MFIF module establishes an \textit{implicit pixel-level spatial alignment} paradigm. Specifically, the LR frames are first independently encoded into a deep feature space via a shared-weight convolutional encoder ($MFIF_{encode}$). A multi-frame fusion operator ($MFIF_{fusion}$) then aggregates these representations, effectively mining sub-pixel complementary information across frames to recover missing high-frequency cues. Finally, a decoder ($MFIF_{decode}$) leveraging sub-pixel convolutions~\cite{shi2016real} named PixelShuffle Block, expands the spatial dimensions of the fused features, naturally aligning them with the HR PAN image without relying on fragile interpolation:
\begin{equation}
\mathbf{F}_{MFIF} = MFIF_{decode}\Big(MFIF_{fusion}\big(\{MFIF_{encode}(\mathbf{I}_{MS, i}^{LR})\}_{i=1}^k\big)\Big),
\end{equation}
where $\mathbf{F}_{MFIF} \in \mathbb{R}^{\gamma H \times \gamma W \times C}$ denotes the HR semantic feature map. Crucially, rather than presenting a rigid concatenation, \textit{SatFusion} acts as a versatile meta-architecture. Any state-of-the-art multi-frame fusion modules  (Fig.~\ref{fig:3}(b)) existing in MFSR (e.g., 2D/3D-CNNs or Transformers~\cite{cornebise2022open,deudon2020highres,molini2019deepsum,salvetti2020multi,an2022tr}) can be elegantly instantiated as the $MFIF_{fusion}$ operator.

\textbf{2) Multi-Source Image Fusion (MSIF):} 
Building upon the implicitly aligned HR semantic features, the MSIF module is dedicated to injecting fine-grained spatial textures from the PAN image. This process yields the detail-rich, multi-source feature map $\mathbf{F}_{MSIF} \in \mathbb{R}^{\gamma H \times \gamma W \times C}$:
\begin{equation}
\mathbf{F}_{MSIF} = MSIF_{fusion}(\mathbf{F}_{MFIF}, \mathbf{I}_{PAN}^{HR}).
\end{equation}
Following the modular philosophy of MFIF, the multi-source fusion module (Fig.~\ref{fig:3}(c)) in Pansharpening~\cite{masi2016Pansharpening,deng2020detail,peng2023u2net,he2025pan,wang2025adaptive} can be seamlessly adopted as the $MSIF_{fusion}$ operator, freeing them from the burden of explicit MS upsampling.

\textbf{3) Fusion Composition:} 
Finally, to adaptively integrate the outputs of MFIF and MSIF, we first aggregate their complementary features via an initial element-wise addition. We then refine this combined representation using a residual convolution block ($ConvBlock$) of stacked $1\times 1$ convolutions, performing content-aware, pixel-wise spectral re-weighting:
\begin{equation}
\mathbf{I}_{MS}^{HR} = ConvBlock(\mathbf{F}_{MFIF} + \mathbf{F}_{MSIF}) + (\mathbf{F}_{MFIF} + \mathbf{F}_{MSIF}).
\end{equation}

\begin{figure}
    \centering
    \includegraphics[width=\linewidth]{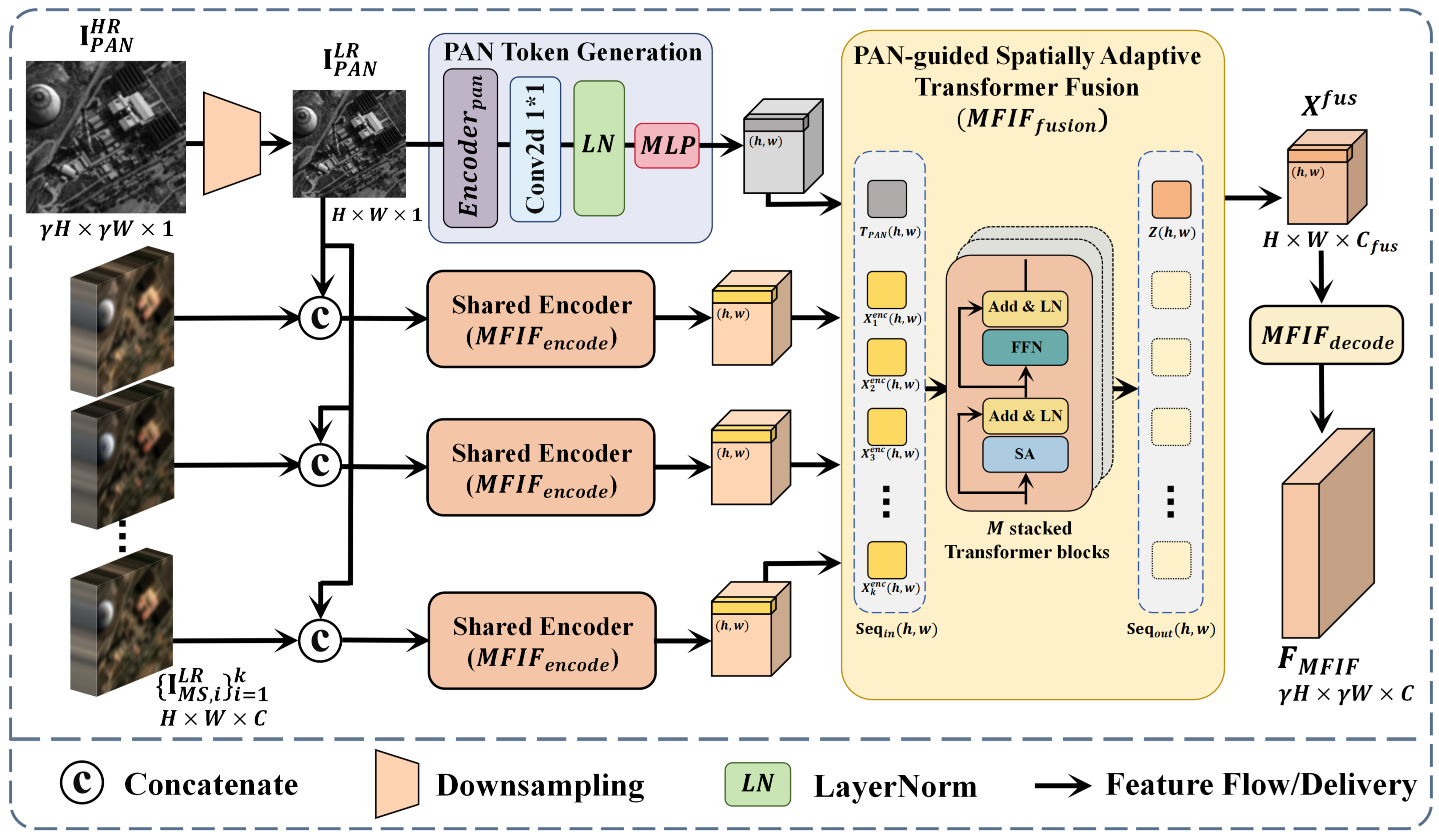}
    \caption{
        Architecture of the advanced MFIF module in \textit{SatFusion*}. It leverages downsampled PAN features as a structural anchor during encoding, and introduces PAN-guided, spatially adaptive tokens for adaptively multi-frame Transformer fusion.
    }
    \label{fig:4}
\end{figure}

\subsection{\textit{SatFusion*}: PAN-Guided Adaptive Aggregation}
\label{sec:3.3}
While \textit{SatFusion} provides a unified framework, its MFIF module still faces two critical limitations. First, it aggregates multi-frame features without explicit guidance from stable spatial structures, leading to suboptimal fusion under severe local misalignment and noise perturbations. Second, existing MFSR methods generally pay limited attention to generalization with respect to the number of input frames, severely restricting their generalization when the number of available frames during inference differs from training. 
As shown in Fig.~\ref{fig:4}, to address these issues, we propose \textit{SatFusion*}, which significantly enhances the MFIF module by optimizing both the encoding ($MFIF_{encode}$) and fusion ($MFIF_{fusion}$) stages through directly injecting PAN guidance.

\textbf{PAN-Guided Encoding:} To alleviate the local misalignment and noise among MS frames, we introduce the downsampled PAN image $\mathbf{I}_{PAN}^{LR} \in \mathbb{R}^{H \times W \times 1}$ as a \emph{weak geometric anchor} during the encoding stage. It is concatenated along the channel dimension with each MS frame prior to the shared-weight encoder:
\begin{equation}
\mathbf{X}_i^{enc} = MFIF_{encode}\big([\mathbf{I}_{MS, i}^{LR}, \mathbf{I}_{PAN}^{LR}]\big).
\end{equation}

\textbf{PAN-Guided Spatially Adaptive Token:} 
To achieve robust generalization across varying frame counts, we implement $MFIF_{fusion}$ using a Transformer architecture, which naturally supports variable-length inputs via self-attention mechanisms~\cite{vaswani2017attention, dosovitskiy2020image,he2022masked,liu2021swin,liu2022swin,li2022blip}. However, standard Transformer-based MFSR methods (e.g., TR-MISR~\cite{an2022tr}) aggregate multi-frame features into a single, globally shared learnable embedding (\emph{CLS token}). A global token fails to capture the rich, spatially varying geometries inherent in RS imagery. 

Unlike previous works, \textit{SatFusion*} introduces \emph{position-specific fusion tokens} dynamically generated from the local PAN structural priors. Specifically, we project the downsampled PAN features into dedicated embedding tokens at each spatial location $(h,w)$:
\begin{equation}
\mathbf{T}_{PAN}(h,w) = MLP\Big(LN\big(Conv_{1\times1}(Encoder_{pan}(\mathbf{I}_{PAN}^{LR}))\big)\Big),
\end{equation}
where $\mathbf{T}_{PAN}(h,w)$ serves as the structural condition. During fusion, at each spatial location $(h,w)$, the encoded features from all $k$ frames $\{\mathbf{X}_i^{enc}(h,w)\}_{i=1}^k$ and the corresponding PAN token form the input sequence for the Transformer encoder blocks ($\mathcal{T}$):
\begin{equation}
\mathbf{Seq}_{out}(h,w) = \mathcal{T}^{(M)} \Big(\big[\mathbf{T}_{PAN}(h,w), \mathbf{X}_1^{enc}(h,w), \dots, \mathbf{X}_k^{enc}(h,w)\big]\Big).
\end{equation}
Specifically, we extract the output vector corresponding to the $\mathbf{T}_{PAN}(h,w)$ token index (position 0) as the local fused representation $\mathbf{Z}(h,w)$.
By performing this extraction in parallel across all spatial locations $(h,w)$, we assemble the final feature map $\mathbf{X}^{fus} = \{\mathbf{Z}(h,w)\}_{h=1,w=1}^{H,W}$, which is subsequently passed to the $MFIF_{decode}$ module for resolution reconstruction. By doing so, the PAN image acts as an active, spatially adaptive query that guides the multi-frame aggregation, seamlessly handling arbitrary frame counts while boosting robustness against input degradation. Detailed tensor dimensionalities and mathematical formulations for this enhanced MFIF module are deferred to Appendix~\ref{app:architecture_3}.

\begin{table*}[t]
\centering
\caption{Quantitative comparison on the WorldStrat Real (a) and Simulated (b) datasets (answering RQ1). For \textit{SatFusion} and \textit{SatFusion*}, we report the average performance across all modular combinations to demonstrate systematic superiority. Detailed results for every individual combination are provided in Appendix~\ref{app:full_results}.}
\label{tab:1}
\small
\renewcommand{\arraystretch}{0.9} 
\setlength{\tabcolsep}{4.5pt}

\definecolor{decline}{RGB}{214,39,40}   
\definecolor{improve}{RGB}{44,160,44}   

\newcommand{\bad}[1]{\ensuremath{_{\scriptscriptstyle \color{decline} #1}}}    
\newcommand{\good}[1]{\ensuremath{_{\scriptscriptstyle \color{improve} #1}}}   
\newcommand{\base}[1]{\ensuremath{_{\scriptscriptstyle \color{gray} #1}}}      

\resizebox{\textwidth}{!}{
\begin{tabular}{l cccccc cccccc} 
\toprule
\multirow{2}{*}{\textbf{Methods}} & \multicolumn{6}{c}{\textbf{(a) Metrics on the WorldStrat Real Dataset}} & \multicolumn{6}{c}{\textbf{(b) Metrics on the WorldStrat Simulated Dataset}} \\
\cmidrule(lr){2-7}\cmidrule(lr){8-13}
& \textbf{PSNR}$\uparrow$ & \textbf{SSIM}$\uparrow$ & \textbf{SAM}$\downarrow$ & \textbf{ERGAS}$\downarrow$ & \textbf{MAE}$\downarrow$ & \textbf{MSE}$\downarrow$ & \textbf{PSNR}$\uparrow$ & \textbf{SSIM}$\uparrow$ & \textbf{SAM}$\downarrow$ & \textbf{ERGAS}$\downarrow$ & \textbf{MAE}$\downarrow$ & \textbf{MSE}$\downarrow$ \\ 
\midrule

\multicolumn{13}{c}{\cellcolor{gray!15}\textbf{(a) MFSR: Fusing Multi-Frame Information}} \\
\midrule
MF-SRCNN~\cite{cornebise2022open} & 36.8263 & 0.8767 & 2.6776 & 9.3946 & 1.4681 & 9.6654 & 39.0772 & 0.8964 & 2.4051 & 5.5574 & 1.0371 & 4.0053 \\
HighRes-Net~\cite{deudon2020highres} & 37.0815 & 0.8763 & 2.2503 & 9.1406 & 1.4498 & 9.5003 & 39.7523 & 0.9025 & 1.6448 & 5.1644 & 0.9519 & 3.4906 \\
RAMS~\cite{salvetti2020multi} & 37.1946 & 0.8793 & 2.3063 & 8.8203 & 1.4097 & 9.1261 & 40.3275 & 0.9101 & 1.4961 & 4.7717 & 0.8771 & 3.0101 \\
TR-MISR~\cite{an2022tr} & 37.0014 & 0.8778 & 2.3378 & 9.2222 & 1.4283 & 9.2299 & 39.5560 & 0.9005 & 1.7136 & 5.2308 & 0.9714 & 3.6965 \\ \cmidrule(lr){1-13}
\textit{Average} & 37.03\bad{\downarrow 10.35} & 0.8775\bad{\downarrow 0.1105} & 2.39\bad{\uparrow 0.48} & 9.14\bad{\uparrow 6.81} & 1.44\bad{\uparrow 1.05} & 9.38\bad{\uparrow 8.95} & 39.68\bad{\downarrow 9.68} & 0.9024\bad{\downarrow 0.0904} & 1.81\bad{\uparrow 0.11} & 5.18\bad{\uparrow 3.46} & 0.96\bad{\uparrow 0.66} & 3.55\bad{\uparrow 3.28} \\ 
\midrule

\multicolumn{13}{c}{\cellcolor{gray!15}\textbf{(b) Pansharpening: Fusing Multi-Source Information}} \\
\midrule
PNN~\cite{masi2016Pansharpening} & 46.4287 & 0.9877 & 2.1886 & 2.6345 & 0.4341 & 0.5246 & 47.5420 & 0.9886 & 1.9289 & 2.3656 & 0.4192 & 0.4289 \\
PanNet~\cite{yang2017pannet} & 45.6398 & 0.9843 & 2.3978 & 3.0284 & 0.4836 & 0.6414 & 48.0819 & 0.9900 & 1.8248 & 1.9513 & 0.3376 & 0.3108 \\
U2Net~\cite{peng2023u2net} & 46.8601 & 0.9859 & 2.2141 & 2.6720 & 0.4182 & 0.4750 & 47.4352 & 0.9910 & 1.9393 & 2.1106 & 0.3707 & 0.3519 \\
Pan-Mamba~\cite{he2025pan} & 45.7723 & 0.9861 & 2.4807 & 2.8916 & 0.4606 & 0.5840 & 47.9158 & 0.9882 & 1.7338 & 2.0379 & 0.3451 & 0.3324 \\
ARConv~\cite{wang2025adaptive} & 46.6602 & 0.9864 & 2.2151 & 2.7638 & 0.4275 & 0.4944 & 46.8972 & 0.9873 & 2.0434 & 2.2976 & 0.3976 & 0.3897 \\ \cmidrule(lr){1-13}
\textit{Average} & 46.27\bad{\downarrow 1.11} & 0.9861\bad{\downarrow 0.0019} & 2.30\bad{\uparrow 0.39} & 2.80\bad{\uparrow 0.47} & 0.44\bad{\uparrow 0.05} & 0.54\bad{\uparrow 0.11} & 47.57\bad{\downarrow 1.79} & 0.9890\bad{\downarrow 0.0038} & 1.89\bad{\uparrow 0.19} & 2.15\bad{\uparrow 0.43} & 0.37\bad{\uparrow 0.07} & 0.36\bad{\uparrow 0.09} \\ 
\midrule

\multicolumn{13}{c}{\cellcolor{blue!10}\textbf{(c) Ours: Fusing Multi-Frame and Multi-Source Information}} \\
\midrule
\rowcolor{blue!5} \textbf{\textit{SatFusion} (Avg.)} & \underline{47.04}\bad{\downarrow 0.34} & \textbf{0.9896}\good{\uparrow 0.0016} & \underline{2.05}\bad{\uparrow 0.14} & \underline{2.51}\bad{\uparrow 0.18} & \underline{0.40}\bad{\uparrow 0.01} & \underline{0.46}\bad{\uparrow 0.03} & \underline{48.89}\bad{\downarrow 0.47} & \underline{0.9924}\bad{\downarrow 0.0004} & \underline{1.76}\bad{\uparrow 0.06} & \underline{1.84}\bad{\uparrow 0.12} & \underline{0.32}\bad{\uparrow 0.02} & \underline{0.29}\bad{\uparrow 0.02} \\ 
\rowcolor{blue!5} \textbf{\textit{SatFusion*} (Avg.)} & \textbf{47.38}\base{\pm 0.00} & \underline{0.9880}\base{\pm 0.0000} & \textbf{1.91}\base{\pm 0.00} & \textbf{2.33}\base{\pm 0.00} & \textbf{0.39}\base{\pm 0.00} & \textbf{0.43}\base{\pm 0.00} & \textbf{49.36}\base{\pm 0.00} & \textbf{0.9928}\base{\pm 0.0000} & \textbf{1.70}\base{\pm 0.00} & \textbf{1.72}\base{\pm 0.00} & \textbf{0.30}\base{\pm 0.00} & \textbf{0.27}\base{\pm 0.00} \\ 
\bottomrule
\end{tabular}
}
\par\smallskip
\raggedright \small \textbf{Bold}/\underline{Underline}: Best/2nd best group avg. \textcolor{decline}{$\downarrow/\uparrow$} / \textcolor{improve}{$\downarrow/\uparrow$}: Worse/better relative to \textit{SatFusion*} (Avg.), which serves as the $\pm 0.00$ reference.
\end{table*}

\subsection{Loss Function}
\label{sec:3.4}
To jointly optimize spatial texture fidelity and spectral consistency, the entire framework is trained end-to-end using a weighted combination of multiple criteria:
\begin{equation}
\label{eq:loss}
\mathcal{L} = \lambda_1 \mathcal{L}_{MAE} + \lambda_2 \mathcal{L}_{MSE} + \lambda_3 \mathcal{L}_{SSIM} + \lambda_4 \mathcal{L}_{SAM},
\end{equation}
where $\mathcal{L}_{MAE}$ and $\mathcal{L}_{MSE}$ enforce pixel-wise reconstruction constraints, $\mathcal{L}_{SSIM}$ maximizes structural similarity for high-frequency details~\cite{wang2004image}, and $\mathcal{L}_{SAM}$ (Spectral Angle Mapper) explicitly mitigates spectral distortions introduced during multi-source integration~\cite{li2018single}. $\lambda_1$ to $\lambda_4$ are balancing hyper-parameters.

\section{Experiments and Results}
\label{sec:4}
\subsection{Datasets}
\label{sec:4.1}
To comprehensively evaluate the proposed framework, we conduct experiments on both real-world and simulated satellite datasets.

\textbf{Real-World Dataset:} The WorldStrat dataset~\cite{cornebise2022open} provides real-world, multi-frame LR MS images paired with temporally matched HR PAN and MS images from SPOT~6/7. By natively retaining degraded observations rather than artificially filtering them out, it serves as a highly representative benchmark for practical satellite imaging conditions. 

\textbf{Simulated Datasets:} Standard Pansharpening datasets~\cite{deng2022machine} typically provide single-frame LR MS and HR PAN pairs generated via the ideal Wald protocol~\cite{wald1997fusion}, which assumes clean inputs and enforces perfect pixel-level alignment. To rigorously evaluate model robustness under practical satellite imaging conditions, we introduce a physics-inspired image formation strategy on the WV3, QB, and GF2 datasets. Specifically, we explicitly model sub-pixel spatial misalignment, sensor PSF/MTF blurring, and mixed sensor noise to generate realistic, degraded multi-frame sequences ($\{\mathbf{I}_{MS,i}^{LR}\}_{i=1}^{k}$). This approach effectively bridges the gap between ideal simulations and practical satellite imaging conditions. Detailed synthesis procedures, including the simulation algorithm and comparative visualization, are provided in Appendix~\ref{app:dataset}.

\subsection{Training and Evaluation}
\label{sec:4.2}

\subsubsection{Training}
\label{sec:4.2.1}
To ensure a fair comparison, our \textit{SatFusion} variants and all baseline methods are trained from scratch under strictly identical experimental settings (e.g., matching optimizers, batch sizes, and spatial dimensions). All models are optimized on a server equipped with eight NVIDIA RTX 4090 GPUs. The code is available at: \href{https://github.com/yufeiTongZJU/SatFusion.git}{https://github.com/yufeiTongZJU/SatFusion.git}. Detailed hyper-parameter configurations for both the network and training process are deferred to Appendix~\ref{app:settings}.
Furthermore, when training on real-world datasets like WorldStrat, inherent acquisition differences often introduce global brightness variations and sub-pixel spatial shifts between the LR inputs and the HR ground truth (GT)~\cite{bhat2021deep,deudon2020highres,molini2019deepsum,salvetti2020multi,an2022tr,di2025qmambabsr}. Following prior works, on the WorldStrat dataset, we apply global brightness alignment and spatial shift correction to the reconstructed images prior to loss computation. This necessary calibration is also applied before quantitative evaluation during testing, and is consistently enforced across all evaluated methods to guarantee a rigorous and unbiased comparison.

\subsubsection{Evaluation}
\label{sec:4.2.2}
Our extensive experiments are systematically designed to address five core research questions (RQs) using a comprehensive suite of quantitative metrics (PSNR, SSIM, MAE, MSE, SAM~\cite{yuhas1992discrimination}, and ERGAS~\cite{alparone2007comparison}) and qualitative visual assessments:

\textbf{RQ1:} How do \textit{SatFusion} and \textit{SatFusion*} perform against state-of-the-art baselines across both real-world and realistically simulated datasets? 
\textbf{RQ2:} How do the number of input frames $k$ and the super-resolution scale factor $\gamma$ affect model performance? 
\textbf{RQ3:} How well do the models generalize under different levels of noise perturbations and when the number of input frames differs between training and inference? 
\textbf{RQ4:} How do the designs of individual modules and the choice of loss functions influence the performance of \textit{SatFusion} and \textit{SatFusion*}? 
\textbf{RQ5:} Why does unifying multi-frame and multi-source paradigms yield superior reconstruction fidelity compared to isolated MFSR or Pansharpening approaches?

\begin{table*}[htbp]
\centering
\caption{Summarized experimental metrics on the WV3, GF2, and QB simulated datasets. \textit{SatFusion} variants consistently outperform traditional Pansharpening methods. Detailed combinations are provided in Appendix~\ref{app:full_results_wv3}.}
\label{tab:2}
\small
\renewcommand{\arraystretch}{1}
\setlength{\tabcolsep}{4.5pt}

\definecolor{decline}{RGB}{214,39,40}   
\definecolor{improve}{RGB}{44,160,44}   

\newcommand{\bad}[1]{\ensuremath{_{\scriptscriptstyle \color{decline} #1}}}    
\newcommand{\good}[1]{\ensuremath{_{\scriptscriptstyle \color{improve} #1}}}   
\newcommand{\base}[1]{\ensuremath{_{\scriptscriptstyle \color{gray} #1}}}      

\resizebox{\textwidth}{!}{
\begin{tabular}{l cccc cccc cccc}
\toprule
\multirow{2}{*}{\textbf{Methods}} & \multicolumn{4}{c}{\textbf{WV3}} & \multicolumn{4}{c}{\textbf{GF2}} & \multicolumn{4}{c}{\textbf{QB}} \\ 
\cmidrule(lr){2-5} \cmidrule(lr){6-9} \cmidrule(lr){10-13}
& \textbf{PSNR}$\uparrow$ & \textbf{SSIM}$\uparrow$ & \textbf{SAM}$\downarrow$ & \textbf{ERGAS}$\downarrow$ & \textbf{PSNR}$\uparrow$ & \textbf{SSIM}$\uparrow$ & \textbf{SAM}$\downarrow$ & \textbf{ERGAS}$\downarrow$ & \textbf{PSNR}$\uparrow$ & \textbf{SSIM}$\uparrow$ & \textbf{SAM}$\downarrow$ & \textbf{ERGAS}$\downarrow$ \\ 
\midrule

\multicolumn{13}{c}{\cellcolor{gray!15}\textbf{(a) Pansharpening: Fusing Multi-Source Information}} \\
\midrule
PNN~\cite{masi2016Pansharpening} & 36.5340 & 0.9548 & 4.2758 & 3.6505 & 41.9402 & 0.9696 & 1.5319 & 1.4494 & 36.0032 & 0.9264 & 5.1423 & 5.7539 \\
DiCNN~\cite{he2019Pansharpening} & 37.1690 & 0.9611 & 4.0397 & 3.4236 & 42.4487 & 0.9729 & 1.4386 & 1.3750 & 36.2339 & 0.9305 & 4.9892 & 5.6132 \\
MSDCNN~\cite{yuan2018multiscale} & 35.9721 & 0.9454 & 4.7528 & 3.9338 & 42.0847 & 0.9702 & 1.5241 & 1.4278 & 35.8757 & 0.9254 & 5.1286 & 5.8496 \\
DRPNN~\cite{wei2017boosting}     & 37.1089 & 0.9603 & 4.1000 & 3.4253 & 43.1093 & 0.9760 & 1.3330 & 1.2747 & 37.3074 & 0.9436 & 4.7667 & 4.9032 \\
FusionNet~\cite{deng2020detail}  & 37.5678 & 0.9634 & 3.8872 & 3.2372 & 42.7230 & 0.9740 & 1.3562 & 1.3319 & 36.8057 & 0.9379 & 4.9236 & 5.1991 \\
U2Net~\cite{peng2023u2net}       & 38.0416 & 0.9678 & 3.6772 & 3.0081 & 43.1198 & 0.9763 & 1.2491 & 1.2930 & 37.7626 & 0.9479 & 4.6238 & 4.6672 \\
\cmidrule(lr){1-13}
\textit{Average} & 37.07\bad{\downarrow 0.96} & 0.9588\bad{\downarrow 0.0093} & 4.12\bad{\uparrow 0.47} & 3.45\bad{\uparrow 0.34} & 42.57\bad{\downarrow 2.02} & 0.9732\bad{\downarrow 0.0085} & 1.41\bad{\uparrow 0.27} & 1.36\bad{\uparrow 0.26} & 36.66\bad{\downarrow 1.11} & 0.9353\bad{\downarrow 0.0144} & 4.93\bad{\uparrow 0.50} & 5.33\bad{\uparrow 0.66} \\
\midrule

\multicolumn{13}{c}{\cellcolor{blue!10}\textbf{(b) Ours: Fusing Multi-Frame and Multi-Source Information}} \\
\midrule
\rowcolor{blue!5} \textbf{\textit{SatFusion} (Avg.)} & \underline{37.71}\bad{\downarrow 0.32} & \underline{0.9665}\bad{\downarrow 0.0016} & \underline{3.77}\bad{\uparrow 0.12} & \underline{3.20}\bad{\uparrow 0.09} & \underline{43.52}\bad{\downarrow 1.07} & \underline{0.9778}\bad{\downarrow 0.0039} & \underline{1.24}\bad{\uparrow 0.10} & \underline{1.25}\bad{\uparrow 0.15} & \underline{37.47}\bad{\downarrow 0.30} & \underline{0.9466}\bad{\downarrow 0.0031} & \underline{4.53}\bad{\uparrow 0.10} & \underline{4.84}\bad{\uparrow 0.17} \\  
\rowcolor{blue!5} \textbf{\textit{SatFusion*} (Avg.)} & \textbf{38.03}\base{\pm 0.00} & \textbf{0.9681}\base{\pm 0.0000} & \textbf{3.65}\base{\pm 0.00} & \textbf{3.11}\base{\pm 0.00} & \textbf{44.59}\base{\pm 0.00} & \textbf{0.9817}\base{\pm 0.0000} & \textbf{1.14}\base{\pm 0.00} & \textbf{1.10}\base{\pm 0.00} & \textbf{37.77}\base{\pm 0.00} & \textbf{0.9497}\base{\pm 0.0000} & \textbf{4.43}\base{\pm 0.00} & \textbf{4.67}\base{\pm 0.00}  \\
\bottomrule
\end{tabular}
}
\par\smallskip
\raggedright \small \textbf{Bold}/\underline{Underline}: Best/2nd best group avg. \textcolor{decline}{$\downarrow/\uparrow$} / \textcolor{improve}{$\downarrow/\uparrow$}: Worse/better relative to \textit{SatFusion*} (Avg.), which serves as the $\pm 0.00$ reference.
\end{table*}

\subsection{Main Results (RQ1)}
\label{sec:4.3}

\subsubsection{Results on WorldStrat}
\label{sec:4.3.1}
We first evaluate our framework on the WorldStrat dataset, utilizing both the original real-world data and the simulated data constructed via our physics-inspired pipeline. In our experimental setup, MFSR baselines strictly process $k=8$ LR MS frames, and Pansharpening baselines fuse a randomly selected LR MS frame with the HR PAN image. In contrast, our \textit{SatFusion} variants comprehensively leverage both the 8-frame MS sequence and the PAN image. All evaluated methods are trained using the same joint loss function (Eq.~\ref{eq:loss}) with fixed weights ($\lambda_1 = 0.3$, $\lambda_2 = 0.3$, $\lambda_3 = 0.2$, $\lambda_4 = 0.2$) and a spatial upscaling factor of $\gamma = 3$.
As detailed in Sec.~\ref{sec:3.2}, the unified interfaces ($MFIF_{fusion}$ and $MSIF_{fusion}$) enable \textit{SatFusion} to seamlessly integrate diverse multi-frame and multi-source components from existing literature into a cohesive architecture. To present a concise and impactful comparison, Table~\ref{tab:1} reports the average performance of \textit{SatFusion} and \textit{SatFusion*} across all instantiated modular combinations. The exhaustive quantitative results for every specific combination are deferred to Appendix~\ref{app:full_results}.

As demonstrated in Table~\ref{tab:1}, our unified approach fundamentally outperforms isolated paradigms. Compared to the MFSR baseline average, \textit{SatFusion} yields a remarkable performance leap, improving PSNR by \textbf{25.1\%} and reducing ERGAS by \textbf{69.6\%}. This substantial gap validates our core motivation: injecting HR PAN structural priors is essential to shatter the inherent MFSR performance ceiling. Furthermore, compared to the Pansharpening average, \textit{SatFusion} achieves a \textbf{2.2\%} PSNR gain and a \textbf{12.0\%} ERGAS reduction. This demonstrates that leveraging multi-frame complementary information for implicit alignment is far more effective than direct single-frame upsampling. Notably, \textit{SatFusion*} delivers the best overall performance across the evaluation metrics. By jointly modeling multi-frame and multi-source observations within a structurally-guided feature space, \textit{SatFusion*} optimally couples spatial and temporal representations. Qualitative visual comparisons (detailed in Appendix~\ref{app:qual_worldstrat}) further corroborate these findings.

\subsubsection{Results on WV3, QB, and GF2}
\label{sec:4.3.2}
We further evaluate \textit{SatFusion} and \textit{SatFusion*} on the WV3, QB, and GF2 datasets using the physics-inspired simulated data, in order to examine their advantages over Pansharpening. These classical benchmarks have been extensively studied in the Pansharpening literature. Accordingly, we select six highly representative Pansharpening architectures as our baselines. To instantiate \textit{SatFusion}, we integrate MFSR components into the $MFIF_{fusion}$ interface. All baselines are trained using their original configurations (e.g., image settings, epochs) provided by the official DLPan-Toolbox codebase~\cite{deng2022machine}. Our \textit{SatFusion} variants natively inherit these exact training hyperparameters from their corresponding Pansharpening baselines, with modifications restricted purely to the network architecture and our joint loss formulation. Table~\ref{tab:2} presents the average performance of the unified \textit{SatFusion} combinations against the baselines. Readers are referred to Appendix~\ref{app:full_results_wv3} for the exhaustive, instance-by-instance quantitative evaluation.

The results in Table~\ref{tab:2} unequivocally highlight the limitations of traditional explicit alignment when handling inputs. Under complex noise and spatial misalignment, isolated Pansharpening models experience significant performance bottlenecks. In contrast, by effectively leveraging sub-pixel complementary information across multiple frames, \textit{SatFusion} achieves an average PSNR improvement of \textbf{2.7\%} and an average ERGAS reduction of \textbf{9.6\%} across the three datasets compared to the baseline average. Furthermore, \textit{SatFusion*} expands this lead, demonstrating that embedding PAN-guided adaptive priors into the multi-frame modeling process effectively strengthens the deep coupling and implicit alignment of multi-frame and multi-source features. 
Qualitative visual comparisons (provided in Appendix~\ref{app:qual_simulated}) further corroborate these findings.

\section{Analysis}
\label{sec:5}
To provide deeper insights into our unified framework and answer RQ2--RQ5, we conduct a series of detailed analyses. For conciseness in the following experiments, we consistently instantiate \textit{SatFusion} and \textit{SatFusion*} using highly representative backbones (i.e., HighRes-Net and PanNet for the WorldStrat dataset; TR-MISR and FusionNet for the QB dataset) unless otherwise specified.
\subsection{Hyperparameter Study (RQ2)}
\label{sec:5.1}

\begin{figure}[h]
    \centering
    \includegraphics[width=1.0\linewidth]{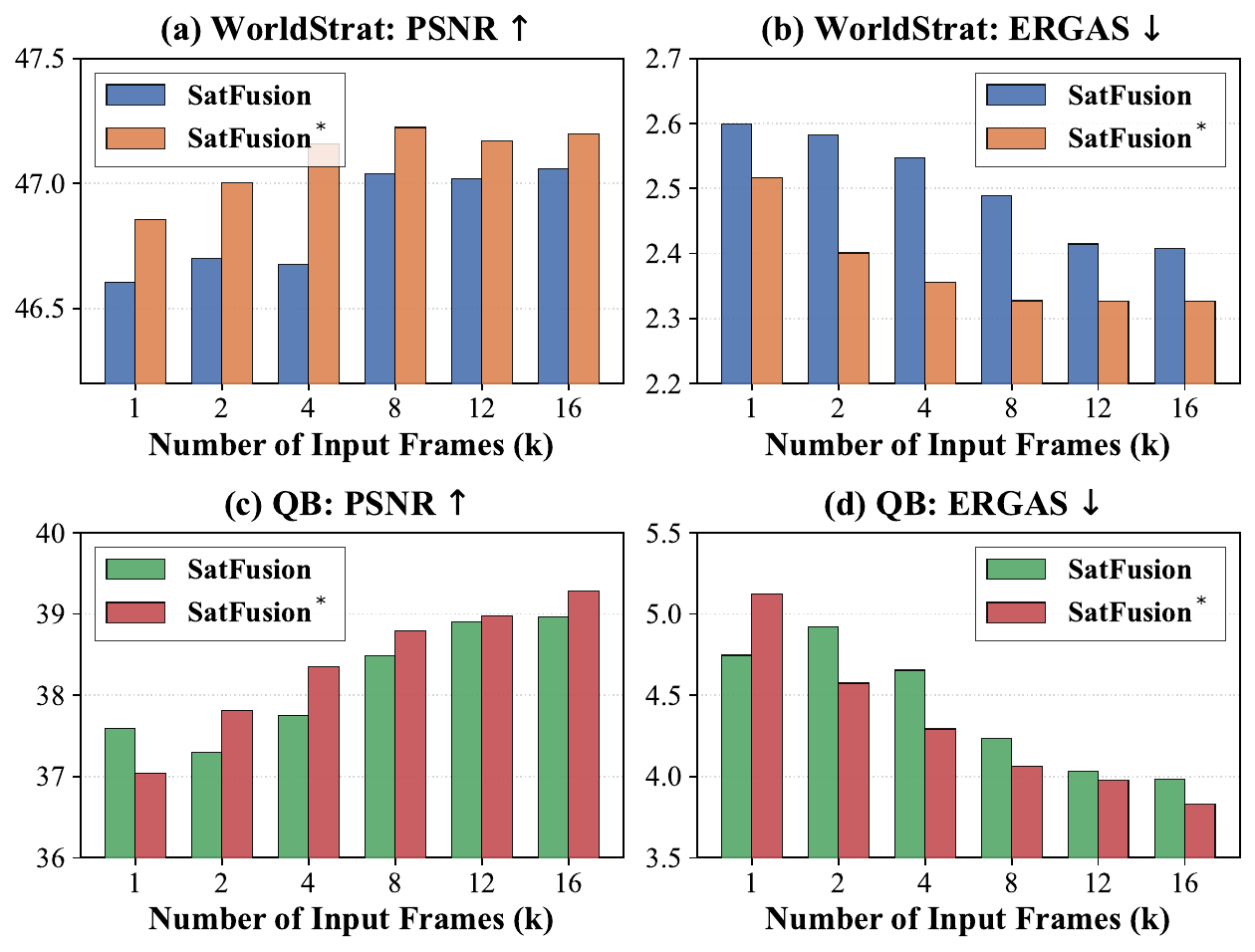}
    \caption{
        Effect of the number of input frames $k$ on fusion performance. We report the PSNR and ERGAS metrics on (a, b) the WorldStrat dataset and (c, d) the QB dataset for both \textit{SatFusion} and \textit{SatFusion*}.
    }
    \label{fig:8}
\end{figure}

\textbf{Effect of Input Image Number:} 
We vary the input sequence length $k$ during training on both the real WorldStrat and simulated QB datasets. As depicted in Fig.~\ref{fig:8}, reconstruction quality exhibits a strict positive correlation with $k$, confirming that our network effectively harvests sub-pixel complementary information across multiple frames. However, performance gains naturally saturate as $k$ grows. This phenomenon is particularly evident on the real-world WorldStrat dataset (where low-quality frames are not artificially filtered), indicating that while additional frames provide more complementary information, they concurrently introduce marginal noise and redundant content that bounds further improvements.

\textbf{Effect of Upscaling Factor:} 
We further evaluate model robustness across varying spatial upscaling factors $\gamma \in \{2, 3, 5\}$ on the WorldStrat dataset. As reported in Table~\ref{tab:3}, larger $\gamma$ values inherently pose more difficult reconstruction challenges, resulting in a general metric degradation across all methods. Nevertheless, both \textit{SatFusion} and \textit{SatFusion*} consistently dominate the isolated baselines at every scale. Notably, even under the extreme $\gamma=5$ setting, where recovering fine-grained details is notoriously difficult, our methods maintain stable and superior reconstruction quality, demonstrating the strong adaptability of our unified framework.

\begin{table}[h]
\centering
\caption{Quantitative results at different upscaling factors $\gamma$ on the WorldStrat dataset.}
\label{tab:3}
\small
\renewcommand{\arraystretch}{0.75}
\setlength{\tabcolsep}{3.5pt}
\resizebox{\columnwidth}{!}{
\begin{tabular}{clcccc} 
\toprule
\textbf{$\gamma$} & \textbf{Method} & \textbf{PSNR}$\uparrow$ & \textbf{SSIM}$\uparrow$ & \textbf{SAM}$\downarrow$ & \scriptsize{\textbf{ERGAS}$\downarrow$} \\ 
\midrule
\multirow{4}{*}{\rotatebox{90}{$\gamma=2$}} 
 & MFSR & 37.4654 & 0.8832 & 2.3514 & 8.1199 \\
 & Pansharpening & 46.1225 & 0.9801 & 2.7920 & 4.0084 \\
 & \textit{SatFusion} & \underline{47.9195} & \textbf{0.9912} & \underline{1.8548} & \underline{2.2721} \\
 & \textit{SatFusion*} & \textbf{48.0910} & \underline{0.9898} & \textbf{1.7931} & \textbf{2.1879} \\ 
\midrule 

\multirow{4}{*}{\rotatebox{90}{$\gamma=3$}} 
 & MFSR & 37.0815 & 0.8763 & 2.2503 & 9.1406 \\
 & Pansharpening & 45.6398 & 0.9843 & 2.3978 & 3.0284 \\
 & \textit{SatFusion} & \underline{47.0376} & \underline{0.9890} & \underline{2.0267} & \underline{2.4888} \\
 & \textit{SatFusion*} & \textbf{47.2238} & \textbf{0.9898} & \textbf{1.9151} & \textbf{2.3275} \\ 
\midrule

\multirow{4}{*}{\rotatebox{90}{$\gamma=5$}} 
 & MFSR & 35.9801 & 0.8605 & 2.4485 & 8.1229 \\
 & Pansharpening & 44.8784 & 0.9805 & 2.7607 & 3.2184 \\
 & \textit{SatFusion} & \underline{45.9070} & \textbf{0.9870} & \textbf{2.1466} & \underline{2.5871} \\
 & \textit{SatFusion*} & \textbf{46.2071} & \underline{0.9855} & \underline{2.1840} & \textbf{2.5509} \\ 
\bottomrule
\end{tabular}
}
\par\smallskip
\raggedright \small \hspace{2em} \textbf{Bold:} Best; \underline{Underline:} Second best.
\end{table}

\subsection{Generalization Evaluation (RQ3)}
\label{sec:5.2}
\begin{figure}[h]
    \centering
    \includegraphics[width=1.0\linewidth]{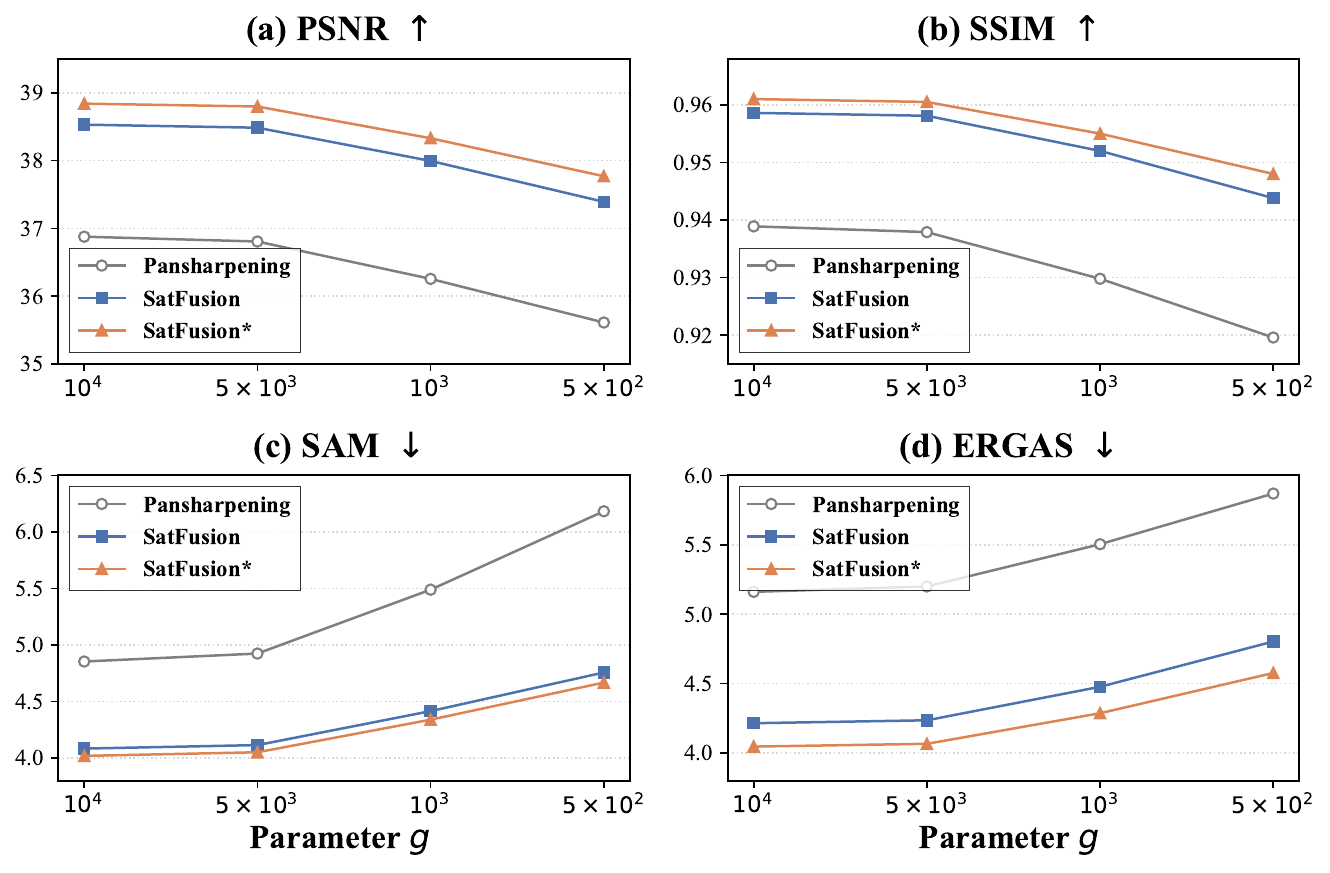}
    \caption{
        Robustness analysis against varying noise intensities (controlled by gain $g$, where smaller is noisier; all models trained at $g=5\times10^3$). Across different noise levels, \textit{SatFusion} and \textit{SatFusion*} consistently outperform Pansharpening methods.  
    }
    \label{fig:9}
\end{figure}

\textbf{Robustness to Image Quality Variations:}
We control the noise intensity during inference by adjusting the photon noise gain $g$ in our physics-inspired simulation pipeline. As shown in Fig.~\ref{fig:9}, both \textit{SatFusion} and \textit{SatFusion*} consistently outperform Pansharpening methods across different noise levels. This demonstrates that exploiting complementary information from multiple frames effectively mitigates noise interference, allowing the proposed framework to retain robust and stable performance in challenging scenarios such as image blur.

\begin{figure}[t]
    \centering
    \includegraphics[width=1.0\linewidth]{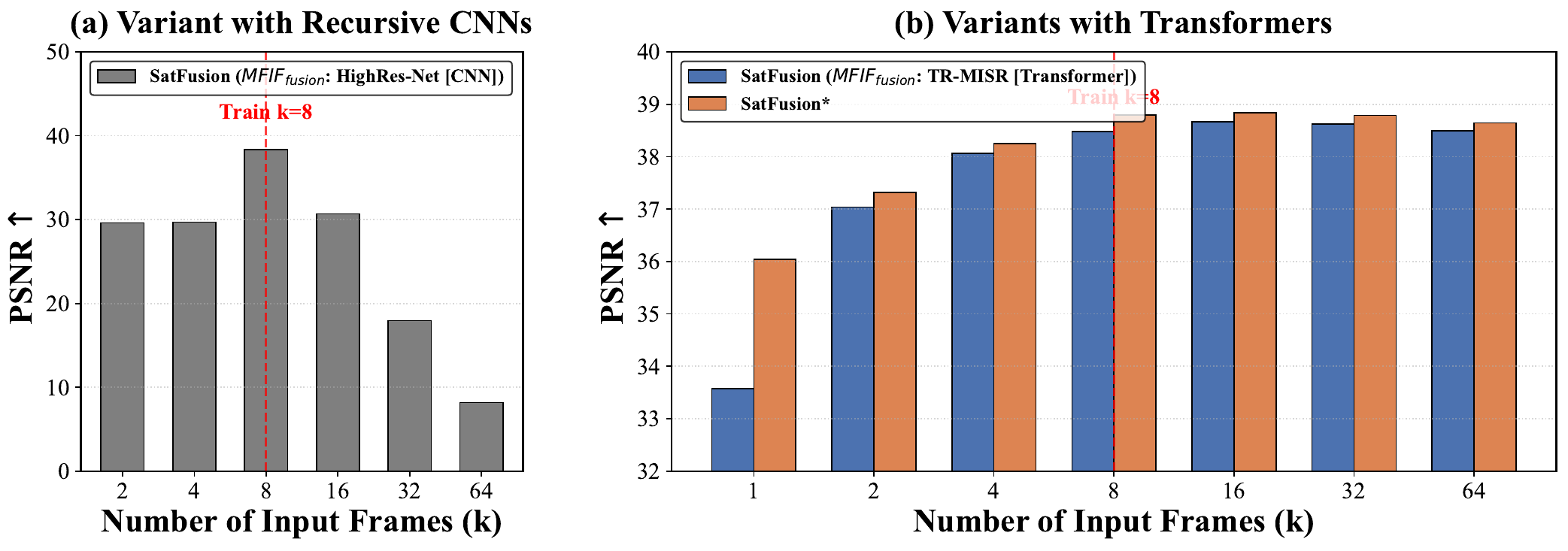}
    \caption{
        Generalization performance under varying inference frame counts $k$. (a) The variant utilizing recursive CNN fusion fails to generalize when tested beyond the training setting. (b) Transformer-based variants maintain stable performance, where \textit{SatFusion*} explicitly demonstrates superior robustness even at extreme sequence lengths.
    }
    \label{fig:10}
\end{figure}

\textbf{Generalization to Inference Frame Counts:}
In real satellite scenarios, the number of available overlapping frames is often highly variable. We evaluate adaptability to variable sequence lengths by modifying the inference frame count $k$ (trained strictly at $k=8$). While concatenation-based methods fail upon length mismatch, recursive CNNs (Fig.~\ref{fig:10}(a)) and Transformers (Fig.~\ref{fig:10}(b)) natively handle variable inputs. As $k$ increases, fusion quality initially improves due to richer complementary information. However, recursive CNN variants exhibit fragile generalization, collapsing when $k$ deviates significantly from the training setting. In contrast, our Transformer-based \textit{SatFusion*} effectively leverages self-attention to filter noise, maintaining peak fidelity and superior stability even at extreme lengths (e.g., $k=64$).

\subsection{Ablation Study (RQ4)}
\label{sec:5.3}

\textbf{Impact of Core Modules:} 
Table~\ref{tab:4} evaluates the necessity of the MFIF, MSIF, and Fusion Composition (FC) modules. When the MFIF module is removed, the framework degrades to single-frame Pansharpening, causing a drastic performance drop due to the loss of complementary information from multiple frames. Conversely, ablating the MSIF module strips away fine-grained PAN textures, severely degrading spatial fidelity. Finally, removing the FC module harms spectral consistency and overall metrics, confirming its essential role as a spectral refinement step. These results confirm that our unified modeling outperforms isolated paradigms.

\begin{table}[h]
\centering
\caption{Ablation of core components on WorldStrat (WS) and QB datasets.}
\label{tab:4}
\renewcommand{\arraystretch}{0.70}
\setlength{\tabcolsep}{3.0pt}
\begin{tabular}{cc ccc cccc} 
\toprule
\textbf{} & \textbf{Data} & \textbf{MFIF} & \textbf{MSIF} & \textbf{FC} & \textbf{PSNR}$\uparrow$ & \textbf{SSIM}$\uparrow$ & \textbf{SAM}$\downarrow$ & \scriptsize{\textbf{ERGAS}$\downarrow$} \\ 
\midrule
\multirow{8}{*}{\rotatebox{90}{\textit{SatFusion}}} & \multirow{4}{*}{WS} 
 & $\times$ & $\checkmark$ & $\checkmark$ & 45.9802 & 0.9847 & 2.2177 & 2.8256 \\
 & & $\checkmark$ & $\times$ & $\checkmark$ & 37.0787 & 0.8758 & 2.2510 & 9.1041 \\
 & & $\checkmark$ & $\checkmark$ & $\times$ & 46.3395 & \textbf{0.9896} & 2.2861 & 2.5255 \\
 & & $\checkmark$ & $\checkmark$ & $\checkmark$ & \textbf{47.0376} & 0.9890 & \textbf{2.0267} & \textbf{2.4888} \\ 
\cmidrule(lr){2-9}
 & \multirow{4}{*}{QB} 
 & $\times$ & $\checkmark$ & $\checkmark$ & 37.0025 & 0.9422 & 4.7216 & 5.0986 \\
 & & $\checkmark$ & $\times$ & $\checkmark$ & 33.3935 & 0.8627 & 5.5241 & 7.8642 \\
 & & $\checkmark$ & $\checkmark$ & $\times$ & 38.3977 & 0.9570 & 4.1362 & 4.2728 \\
 & & $\checkmark$ & $\checkmark$ & $\checkmark$ & \textbf{38.4834} & \textbf{0.9581} & \textbf{4.1139} & \textbf{4.2345} \\ 
\midrule 
\multirow{8}{*}{\rotatebox{90}{\textit{SatFusion*}}} & \multirow{4}{*}{WS} 
 & $\times$ & $\checkmark$ & $\checkmark$ & 46.3021 & 0.9852 & 2.1617 & 2.6767 \\
 & & $\checkmark$ & $\times$ & $\checkmark$ & 40.2804 & 0.9330 & 1.9746 & 4.9765 \\
 & & $\checkmark$ & $\checkmark$ & $\times$ & 46.9979 & 0.9873 & 1.9432 & 2.4328 \\
 & & $\checkmark$ & $\checkmark$ & $\checkmark$ & \textbf{47.2238} & \textbf{0.9898} & \textbf{1.9151} & \textbf{2.3275} \\ 
\cmidrule(lr){2-9}
 & \multirow{4}{*}{QB} 
 & $\times$ & $\checkmark$ & $\checkmark$ & 37.0025 & 0.9422 & 4.7216 & 5.0986 \\
 & & $\checkmark$ & $\times$ & $\checkmark$ & 34.2252 & 0.8881 & 5.2789 & 7.0944 \\
 & & $\checkmark$ & $\checkmark$ & $\times$ & \textbf{38.8061} & 0.9603 & 4.0557 & 4.0723 \\
 & & $\checkmark$ & $\checkmark$ & $\checkmark$ & 38.7960 & \textbf{0.9605} & \textbf{4.0505} & \textbf{4.0650} \\ 
\bottomrule
\end{tabular}
\par\smallskip
\raggedright \small \hspace{2em} \textbf{Bold:} Best; $\checkmark$: w, $\times$: w/o.
\end{table}

\begin{table}[h]
\centering
\caption{Ablation of PAN-guided priors in \textit{SatFusion*}.}
\label{tab:5}
\renewcommand{\arraystretch}{0.75}
\setlength{\tabcolsep}{3.0pt}
\begin{tabular}{cc cc cccc}
\toprule
\textbf{Data} & \textbf{Method} & \textbf{$E_{pan}$} & \textbf{$T_{pan}$} & \textbf{PSNR}$\uparrow$ & \textbf{SSIM}$\uparrow$ & \textbf{SAM}$\downarrow$ & \scriptsize{\textbf{ERGAS}$\downarrow$} \\ 
\midrule
\multirow{4}{*}{WS} & \textit{SatFusion} & $\times$ & $\times$ & 46.7559 & 0.9896 & 2.1167 & 2.5260 \\ \cmidrule(lr){2-8}
 & \multirow{3}{*}{\textit{SatFusion*}} & $\times$ & $\checkmark$ & 47.0941 & 0.9909 & 1.9774 & 2.3694 \\
 & & $\checkmark$ & $\times$ & 47.1625 & \textbf{0.9927} & 1.9191 & 2.3324 \\
 & & $\checkmark$ & $\checkmark$ & \textbf{47.2238} & 0.9898 & \textbf{1.9151} & \textbf{2.3275} \\ 
\midrule 
\multirow{4}{*}{QB} & \textit{SatFusion} & $\times$ & $\times$ & 38.4834 & 0.9581 & 4.1139 & 4.2345 \\ \cmidrule(lr){2-8} 
 & \multirow{3}{*}{\textit{SatFusion*}} & $\times$ & $\checkmark$ & 38.5267 & 0.9586 & 4.1207 & 4.1904 \\
 & & $\checkmark$ & $\times$ & 38.6889 & 0.9597 & 4.0760 & 4.1343 \\
 & & $\checkmark$ & $\checkmark$ & \textbf{38.7960} & \textbf{0.9605} & \textbf{4.0505} & \textbf{4.0650} \\ 
\bottomrule
\end{tabular}
\par\smallskip
\raggedright \small \hspace{2em} \textbf{Bold:} Best; $\checkmark$: w, $\times$: w/o.
\end{table}

\textbf{Effectiveness of PAN-Guided Priors:}
In \textit{SatFusion*}, we intentionally redesigned the MFIF Module to incorporate PAN-guided encoding (denoted as $E_{pan}$) and spatially adaptive tokens (denoted as $T_{pan}$). As shown in Table~\ref{tab:5}, \textit{SatFusion*} outperforms \textit{SatFusion} in fusion quality. Ablating $E_{pan}$ or $T_{pan}$ leads to a drop in performance metrics. This validates that explicitly anchoring multi-frame aggregation with fine-grained, spatially-varying structural priors significantly enhances feature coupling and fusion capability.

\begin{table}[h]
\centering
\caption{Ablation of different loss function configurations.}
\label{tab:6}
\renewcommand{\arraystretch}{0.70}
\setlength{\tabcolsep}{3.0pt}
\begin{tabular}{cc cccc cccc}
\toprule
\textbf{} & \textbf{Data} & 
\shortstack{$\mathcal{L}$ \\[-2pt] \scriptsize{MAE}} & 
\shortstack{$\mathcal{L}$ \\[-2pt] \scriptsize{MSE}} & 
\shortstack{$\mathcal{L}$ \\[-2pt] \scriptsize{SSIM}} & 
\shortstack{$\mathcal{L}$ \\[-2pt] \scriptsize{SAM}} & 
{\textbf{PSNR}$\uparrow$} &
{\textbf{SSIM}$\uparrow$} &
{\textbf{SAM}$\downarrow$} &
{\scriptsize{\textbf{ERGAS}$\downarrow$}}
\\ 
\midrule
\multirow{8}{*}{\rotatebox{90}{\textit{SatFusion}}} & \multirow{4}{*}{WS} 
 & $\checkmark$ & $\checkmark$ & $\checkmark$ & $\checkmark$ & 47.0376 & 0.9890 & 2.0267 & 2.4888 \\
 & & $\checkmark$ & $\checkmark$ & $\checkmark$ & $\times$ & 47.2490 & 0.9903 & \underline{2.2524} & 2.3342 \\
 & & $\checkmark$ & $\checkmark$ & $\times$ & $\checkmark$ & \underline{46.0729} & \underline{0.9878} & 2.1005 & \underline{2.7682} \\
 & & $\times$ & $\checkmark$ & $\times$ & $\times$ & \textbf{43.6225} & \textbf{0.9735} & \textbf{3.3837} & \textbf{3.4150} \\ \cmidrule(lr){2-10}
 & \multirow{4}{*}{QB} 
 & $\checkmark$ & $\checkmark$ & $\checkmark$ & $\checkmark$ & 38.4834 & 0.9581 & 4.1139 & 4.2345 \\
 & & $\checkmark$ & $\checkmark$ & $\checkmark$ & $\times$ & \underline{38.4271} & 0.9577 & \underline{4.1654} & 4.2483 \\
 & & $\checkmark$ & $\checkmark$ & $\times$ & $\checkmark$ & 38.4463 & \underline{0.9573} & 4.0887 & \underline{4.2650} \\
 & & $\times$ & $\checkmark$ & $\times$ & $\times$ & \textbf{38.3706} & \textbf{0.9562} & \textbf{4.2233} & \textbf{4.2926} \\ 
\midrule
\multirow{8}{*}{\rotatebox{90}{\textit{SatFusion*}}} & \multirow{4}{*}{WS} 
 & $\checkmark$ & $\checkmark$ & $\checkmark$ & $\checkmark$ & 47.2238 & 0.9898 & 1.9151 & 2.3275 \\
 & & $\checkmark$ & $\checkmark$ & $\checkmark$ & $\times$ & 47.4381 & 0.9901 & \underline{1.9416} & 2.2670 \\
 & & $\checkmark$ & $\checkmark$ & $\times$ & $\checkmark$ & \underline{46.4639} & \underline{0.9879} & 1.8993 & \underline{2.6645} \\
 & & $\times$ & $\checkmark$ & $\times$ & $\times$ & \textbf{43.6317} & \textbf{0.9804} & \textbf{3.3028} & \textbf{3.3784} \\ \cmidrule(lr){2-10}
 & \multirow{4}{*}{QB} 
 & $\checkmark$ & $\checkmark$ & $\checkmark$ & $\checkmark$ & 38.7960 & 0.9605 & 4.0505 & 4.0650 \\
 & & $\checkmark$ & $\checkmark$ & $\checkmark$ & $\times$ & \underline{38.7492} & 0.9603 & \underline{4.0702} & 4.0770 \\
 & & $\checkmark$ & $\checkmark$ & $\times$ & $\checkmark$ & 38.7926 & \underline{0.9601} & 4.0058 & \underline{4.0892} \\
 & & $\times$ & $\checkmark$ & $\times$ & $\times$ & \textbf{38.6660} & \textbf{0.9586} & \textbf{4.1283} & \textbf{4.1380} \\ 
\bottomrule
\end{tabular}
\par\smallskip
\raggedright \small \hspace{2em} \textbf{Bold:} Worst; \underline{Underline:} Second worst; $\checkmark$: w, $\times$: w/o.
\end{table}

\textbf{Loss Function Design:}
Table~\ref{tab:6} ablates the individual components of our joint loss objective (Eq.~\ref{eq:loss}). Optimizing solely with pixel-wise losses ($\mathcal{L}_{MSE}$) yields the worst overall performance. Removing structural ($\mathcal{L}_{SSIM}$) or spectral ($\mathcal{L}_{SAM}$) constraints distinctly harms high-frequency details and color consistency, respectively. This confirms that our multi-loss formulation is crucial for balancing texture fidelity and spectral preservation.

\begin{figure}[h]
    \centering
    \includegraphics[width=1.0\linewidth]{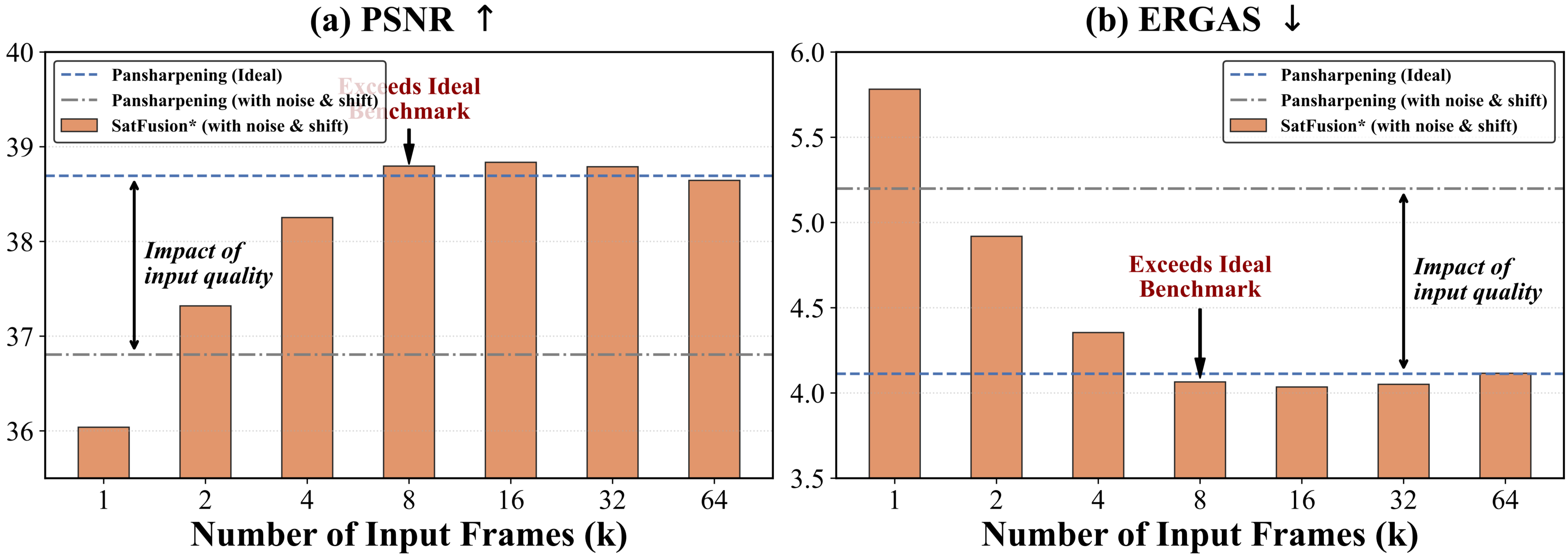}
    \caption{
        Stress test comparison between \textit{SatFusion*} (adverse inputs) and Pansharpening (ideal inputs). By leveraging multi-frame complementary information, \textit{SatFusion*} overcomes severe initial degradation and surpasses even the ideal single-frame benchmark as $k$ increases.
    }
    \label{fig:11}
\end{figure}

\subsection{Advantages over Pansharpening (RQ5)}
\label{sec:5.4}
While our framework's superiority over MFSR intuitively stems from the injection of HR PAN textures, its advantage over Pansharpening requires deeper analysis. To fundamentally answer RQ5, we design an extreme \textbf{stress test} on the QB dataset.

Specifically, we provide the isolated Pansharpening baseline with \textit{ideal, clean inputs} (adhering to the traditional Wald protocol). In stark contrast, we deliberately feed \textit{SatFusion*} with \textit{degraded inputs} (spatial misalignment and compound noise). As illustrated in Fig.~\ref{fig:11}, traditional single-frame Pansharpening is highly sensitive to input quality. However, despite operating at a massive initial disadvantage, \textit{SatFusion*} effectively harvests multi-frame complementary information. Remarkably, as the number of input frames $k$ increases, our method mitigates the severe degradation and eventually surpasses the ideal-case Pansharpening benchmark. This compelling result proves that our unified modeling fundamentally breaks the performance ceiling of traditional isolated fusion paradigms.

\section{Conclusion}
\label{sec:6}
In this work, we present \textit{SatFusion}, a unified framework that fundamentally breaks the isolated paradigms of MFSR and Pansharpening. By jointly fusing multi-frame and multi-source features, \textit{SatFusion} incorporates high-resolution structural priors and circumvents the fragile interpolation bottleneck, while acting as a versatile meta-architecture for existing modules. Furthermore, we introduce \textit{SatFusion*}, which leverages PAN-guided spatially adaptive tokens to robustly handle misalignments and arbitrary frame counts. Extensive evaluations across four diverse datasets demonstrate their effectiveness and practical value in complex RS scenarios.
Moving forward, we plan to explore faithful sensor-aware degradation modeling, broader cross-domain generalization, and scalable efficient inference to tackle extreme geometric misalignments and cross-sensor discrepancies.

\bibliographystyle{ACM-Reference-Format}
\bibliography{refs}

\clearpage 
\appendix

\section*{Appendix Overview}
This appendix provides supplementary technical details for the main paper, including:
\begin{itemize}
    \item Appendix~\ref{app:architecture}: Detailed Network Architecture and Dimensionality.
    \item Appendix~\ref{app:dataset}: Physics-Inspired Dataset Synthesis.
    \item Appendix~\ref{app:settings}: Details of Experimental Parameter Settings.
    \item Appendix~\ref{app:full_results}: Exhaustive Quantitative Results for WorldStrat Modular Combinations.
    \item Appendix~\ref{app:full_results_wv3}: Exhaustive Quantitative Results for WV3, GF2, and QB Modular Combinations.
    \item Appendix~\ref{app:qual_worldstrat}: Qualitative Results on WorldStrat.
    \item Appendix~\ref{app:qual_simulated}: Qualitative Results on WV3, QB, and GF2.
    \item Appendix~\ref{app:implication}: Real-World Implications.
\end{itemize}

\section{Detailed Network Architecture and Dimensionality}
\label{app:architecture}
In this appendix, we provide the detailed dimensional transformations and mathematical formulations for the components within the \textit{SatFusion} and \textit{SatFusion*} frameworks.

\subsection{\textit{SatFusion}: MFIF Module Details}
Given a sequence of $k$ LR MS images $\{\mathbf{I}_{MS,i}^{LR}\}_{i=1}^{k}$, where each $\mathbf{I}_{MS,i}^{LR} \in \mathbb{R}^{H \times W \times C}$, the shared-weight convolutional encoder $MFIF_{encode}$ independently maps them into a deep feature space:
\begin{equation}
\{\mathbf{X}_i^{enc}\}_{i=1}^{k} = MFIF_{encode}\big(\{\mathbf{I}_{MS,i}^{LR}\}_{i=1}^{k}\big),
\end{equation}
where the encoded features $\{\mathbf{X}_i^{enc}\}_{i=1}^{k} \in \mathbb{R}^{k \times H \times W \times C_{enc}}$. 

Subsequently, the fusion operator $MFIF_{fusion}$ aggregates these features along the temporal dimension to form a single, robust feature map:
\begin{equation}
\mathbf{X}^{fus} = MFIF_{fusion}\big(\{\mathbf{X}_i^{enc}\}_{i=1}^{k}\big),
\end{equation}
where $\mathbf{X}^{fus} \in \mathbb{R}^{H \times W \times C_{fus}}$. 

To achieve implicit alignment with the HR PAN image $\mathbf{I}_{PAN}^{HR} \in \mathbb{R}^{\gamma H \times \gamma W \times 1}$, the decoder $MFIF_{decode}$ employs a sub-pixel convolution~\cite{shi2016real} block (PixelShuffle). The feature maps are first passed through a $Conv2d$ layer to adjust the channel dimension to be divisible by $\gamma'^2$, followed by spatial rearrangement:
\begin{equation}
\mathbf{X}^{dec} = PSBlock\big(Conv2d(\mathbf{X}^{fus}), \gamma'\big),
\end{equation}
where $\mathbf{X}^{dec} \in \mathbb{R}^{\gamma' H \times \gamma' W \times C}$, and $\gamma'$ denotes the spatial upscaling factor of the sub-pixel convolution. 

When the structural upscaling factor $\gamma'$ differs from the target task resolution $\gamma$, an optional interpolation-based resizing step is applied to guarantee strict spatial alignment with $\mathbf{I}_{PAN}^{HR}$:
\begin{equation}
\mathbf{F}_{MFIF} =
\begin{cases}
\mathbf{X}^{dec}, & \text{if } \gamma' = \gamma, \\
Resize(\mathbf{X}^{dec}), & \text{otherwise}.
\end{cases}
\end{equation}
By default, we set $\gamma' = \gamma$. The resulting $\mathbf{F}_{MFIF} \in \mathbb{R}^{\gamma H \times \gamma W \times C}$ represents the HR semantic feature map.

\subsection{\textit{SatFusion}: MSIF and Fusion Composition Module Details}
The multi-source fusion component ($MSIF_{fusion}$) integrates the fine-grained texture features of the PAN image into the multi-frame semantic representation, formulated as:
\begin{equation}
\mathbf{F}_{MSIF} = MSIF_{fusion}(\mathbf{F}_{MFIF}, \mathbf{I}_{PAN}^{HR}),
\end{equation}
yielding the detail-enhanced feature map $\mathbf{F}_{MSIF} \in \mathbb{R}^{\gamma H \times \gamma W \times C}$.

Finally, the Fusion Composition module performs residual integration. We first construct an intermediate residual representation $\mathbf{X}^{res}$:
\begin{equation}
\mathbf{X}^{res} = \mathbf{F}_{MFIF} + \mathbf{F}_{MSIF}.
\end{equation}
Then, a sequence of $1 \times 1$ convolutions ($ConvBlock$) applies content-adaptive, pixel-wise spectral re-weighting to refine the fusion outcome:
\begin{equation}
\mathbf{I}_{MS}^{HR} = ConvBlock(\mathbf{X}^{res}) + \mathbf{X}^{res},
\end{equation}
producing the final high-resolution MS image $\mathbf{I}_{MS}^{HR} \in \mathbb{R}^{\gamma H \times \gamma W \times C}$.

\subsection{\textit{SatFusion*}: Enhanced MFIF Details}
\label{app:architecture_3}
In \textit{SatFusion*}, the MFIF module is optimized by introducing PAN guidance into both the encoding and fusion stages. First, the PAN image is downsampled to match the spatial resolution of the LR MS inputs:
\begin{equation}
\mathbf{I}_{PAN}^{LR} = \mathrm{Downsampling}(\mathbf{I}_{PAN}^{HR}).
\end{equation}
During encoding, $\mathbf{I}_{PAN}^{LR} \in \mathbb{R}^{H \times W \times 1}$ is concatenated with each MS frame along the channel dimension:
\begin{equation}
\mathbf{X}_i^{enc} = MFIF_{encode}\big([\mathbf{I}_{MS, i}^{LR}, \mathbf{I}_{PAN}^{LR}]\big).
\end{equation}
where $\{\mathbf{X}_i^{enc}\}_{i=1}^{k} \in \mathbb{R}^{k \times H \times W \times C_{enc}}$.

To generate the spatially adaptive tokens, $\mathbf{I}_{PAN}^{LR}$ is passed through a dedicated PAN encoder:
\begin{equation}
\mathbf{Y}^{enc} = Encoder_{pan}(\mathbf{I}_{PAN}^{LR}),
\end{equation}
where $\mathbf{Y}^{enc} \in \mathbb{R}^{H \times W \times C}$. At each spatial location $(h,w)$, the token $\mathbf{T}_{PAN}(h,w) \in \mathbb{R}^{C_{enc}}$ is derived via position-wise mapping:
\begin{equation}
\mathbf{T}_{PAN}(h,w) = MLP\Big(LN\big(Conv_{1\times1}(\mathbf{Y}^{enc}(h,w))\big)\Big).
\end{equation}

During the Transformer-based fusion process, the input sequence at location $(h,w)$ is constructed as:
\begin{equation}
\mathbf{Seq}_{in}(h,w) = \big[\mathbf{T}_{PAN}(h,w), \mathbf{X}_1^{enc}(h,w), \dots, \mathbf{X}_k^{enc}(h,w)\big].
\end{equation}
This sequence is processed by $M$ stacked Transformer blocks:
\begin{equation}
\mathbf{Seq}_{out}(h,w) = \mathcal{T}^{(M)} \circ \mathcal{T}^{(M-1)} \circ \dots \circ \mathcal{T}^{(1)}\big(\mathbf{Seq}_{in}(h,w)\big).
\end{equation}
The fused representation for location $(h,w)$ is extracted from the position corresponding to the PAN token:
\begin{equation}
\mathbf{Z}(h,w) = \mathbf{Seq}_{out}(h,w)\big|_{index=0}.
\end{equation}
By performing this in parallel across all spatial locations, the final fused feature map is obtained:
\begin{equation}
\mathbf{X}^{fus} = \{\mathbf{Z}(h,w)\}_{h=1,w=1}^{H,W},
\end{equation}
where $\mathbf{X}^{fus} \in \mathbb{R}^{H \times W \times C_{fus}}$ is subsequently passed to the $MFIF_{decode}$ module.

\section{Physics-Inspired Dataset Synthesis}
\label{app:dataset}
In real-world satellite imaging, the acquired multi-frame LR MS images inherently suffer from spatial misalignment, blurring, and sensor noise. Traditional Pansharpening benchmarks typically employ the Wald protocol~\cite{wald1997fusion} to construct simulated datasets, which strictly enforces perfect pixel-level alignment and assumes noise-free conditions. To rigorously evaluate the robustness of \textit{SatFusion} and \textit{SatFusion*}, we introduce a physics-inspired image formation strategy to generate realistically degraded multi-frame LR MS sequences from a single HR MS image.

\begin{figure}
    \centering
    \includegraphics[width=0.9\linewidth]{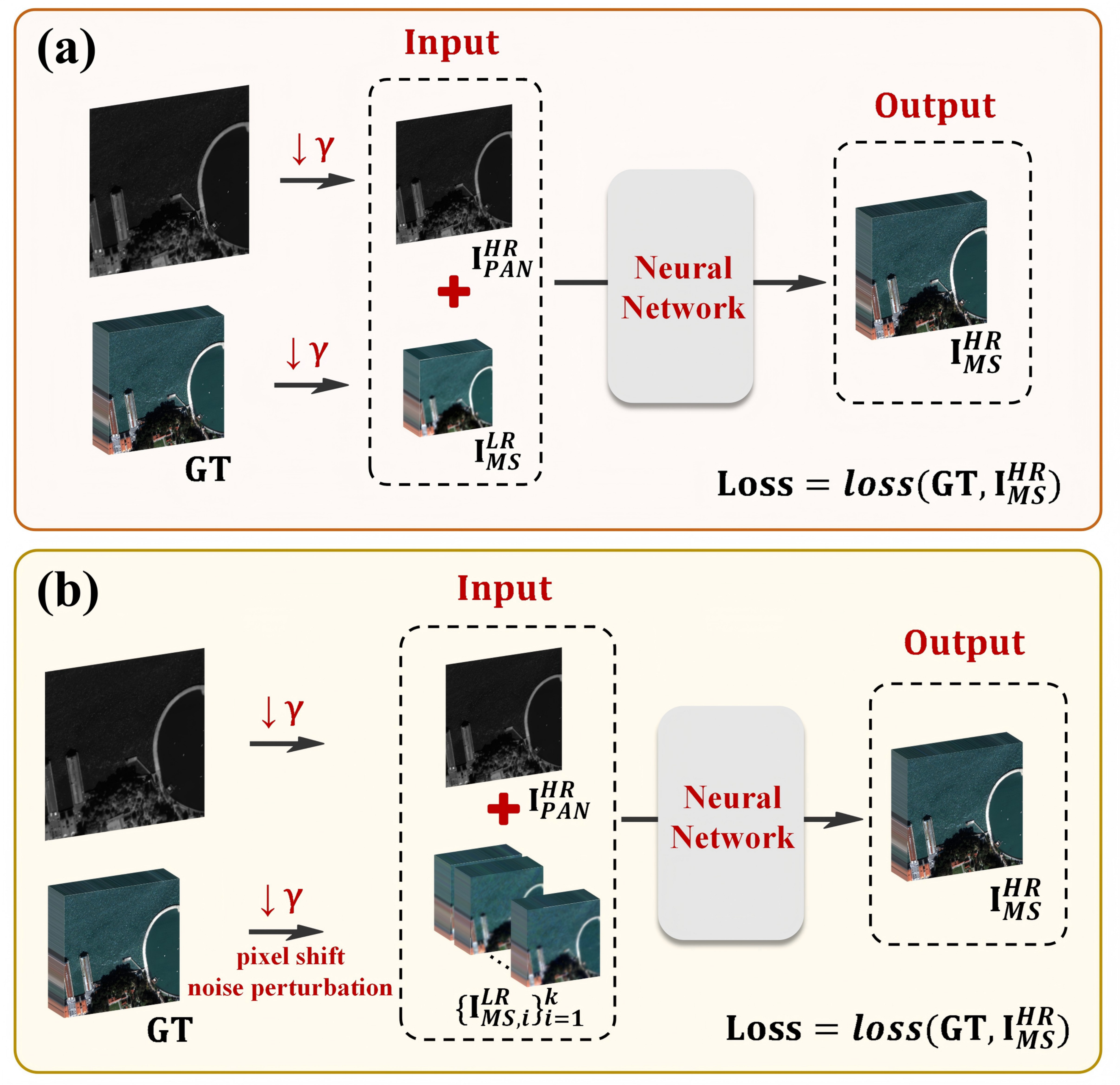}
    \caption{
        Comparison of dataset construction workflows. (a) The standard Wald protocol assumes clean, perfectly aligned inputs. (b) Our proposed physics-inspired synthesis injects sub-pixel misalignment, blur, and mixed noise to simulate realistic satellite imaging degradation.
    }
    \label{fig:5}
\end{figure}

The detailed simulation pipeline is summarized in Algorithm~\ref{alg:synthesis} and conceptually compared with the standard Wald protocol in Fig.~\ref{fig:5}. We explicitly model satellite attitude variations and orbital shifts via random sub-pixel translations. Sensor point spread function (PSF) and modulation transfer function (MTF) effects are approximated by varying-scale Gaussian blur. Following spatial downsampling, we inject both Poisson shot noise and Gaussian readout noise to emulate the complex degradation inherent in practical photon capture processes. 
By applying these diverse degradations, the resulting multi-frame sequence $\{\mathbf{I}_{MS,i}^{LR}\}_{i=1}^{k}$ moves beyond the ideal premise of perfect alignment with the corresponding PAN image, thereby creating a highly challenging testing environment that accurately reflects the complexities of practical satellite imaging conditions.

\begin{algorithm}[h]
\caption{Physics-Inspired $\{\mathbf{I}_{MS,i}^{LR}\}_{i=1}^{k}$ Synthesis}
\label{alg:synthesis}
\begin{algorithmic}[1]
\Require High-resolution MS image $\mathbf{I}_{MS}^{HR} \in \mathbb{R}^{\gamma H \times \gamma W \times C}$;
PSF blur range $\sigma \in [\sigma_{\min}, \sigma_{\max}]$;
sub-pixel shift range $\Delta \in [\Delta_{\min}, \Delta_{\max}]$;
shot noise gain $g$;
readout noise standard deviation $\sigma_r$
\Ensure Multi-frame low-resolution MS set $\{\mathbf{I}_{MS,i}^{LR}\}_{i=1}^{k}$

\For{$i = 1$ to $k$}
    \State Sample sub-pixel shifts $(\delta_x, \delta_y) \sim \mathcal{U}(\Delta)$
    \State $\mathbf{X}_i \leftarrow \text{Warp}(\mathbf{I}_{MS}^{HR}, \delta_x, \delta_y)$
    \Comment{Sub-pixel spatial misalignment}
    
    \State Sample blur scale $\sigma \sim \mathcal{U}([\sigma_{\min}, \sigma_{\max}])$
    \State $\widetilde{\mathbf{X}}_i \leftarrow \text{GaussianBlur}(\mathbf{X}_i, \sigma)$
    \Comment{Sensor PSF / MTF simulation}
    
    \State $\mathbf{I}_{MS,i}^{LR} \leftarrow \text{Downsample}(\widetilde{\mathbf{X}}_i, \gamma)$
    \Comment{Spatial resolution degradation}
    
    \If{$g > 0$}
        \State $\mathbf{I}_{MS,i}^{LR} \leftarrow \mathbf{I}_{MS,i}^{LR} + \mathcal{N}(0, \sqrt{\mathbf{I}_{MS,i}^{LR} / g})$
        \Comment{Shot noise (Gaussian approx.)}
    \EndIf
    
    \State $\mathbf{I}_{MS,i}^{LR} \leftarrow \mathbf{I}_{MS,i}^{LR} + \mathcal{N}(0, \sigma_r^2)$
    \Comment{Readout noise}

\EndFor

\State \Return $\{\mathbf{I}_{MS,i}^{LR}\}_{i=1}^{k}$
\end{algorithmic}
\end{algorithm}

\begin{table}[h]
\centering
\caption{Detailed training hyperparameters for different Pansharpening methods. Our \textit{SatFusion} variants inherit the exact parameters of the respective $MSIF_{fusion}$ backbone employed.}
\label{tab:parameters}
\renewcommand{\arraystretch}{1.0}
\resizebox{\columnwidth}{!}{
\begin{tabular}{lcccccc}
\toprule
\textbf{Hyperparameters} & \textbf{PNN} & \textbf{DiCNN} & \textbf{MSDCNN} & \textbf{DRPNN} & \textbf{FusionNet} & \textbf{U2Net} \\
\midrule
Epochs        & 1000 & 1000 & 500 & 500 & 400 & 400 \\
Batch Size    & 64   & 64 & 64 & 32 & 32 & 32 \\
Optimizer     & SGD  & Adam & Adam & Adam & Adam & Adam \\
Loss Function & $\mathcal{L}_{MSE}$ & $\mathcal{L}_{MSE}$ & $\mathcal{L}_{MSE}$ & $\mathcal{L}_{MSE}$ & $\mathcal{L}_{MSE}$ & $\mathcal{L}_{MSE}$ \\
\bottomrule
\end{tabular}
}
\end{table}

\begin{table*}[h]
\centering
\caption{Exhaustive quantitative metrics on the WorldStrat Real (a) and Simulated (b) datasets. This table details every instantiated combination within the \textit{SatFusion} and \textit{SatFusion*} frameworks, corresponding to the average summaries presented in Table 1 of the main manuscript.}
\label{tab:full_results}
\small
\renewcommand{\arraystretch}{0.75}
\setlength{\tabcolsep}{4pt}

\resizebox{\textwidth}{!}{
\begin{tabular}{ccl cccccc cccccc r} 
\toprule
\multicolumn{3}{c}{\multirow{2}{*}{\textbf{Methods}}} & \multicolumn{6}{c}{\textbf{(a) Metrics on the WorldStrat Real Dataset}} & \multicolumn{6}{c}{\textbf{(b) Metrics on the WorldStrat Simulated Dataset}} & \multirow{2}{*}{\textbf{\#Params}} \\
\cmidrule(lr){4-9}\cmidrule(lr){10-15}
& & &
\multicolumn{1}{c}{\textbf{PSNR}$\uparrow$} &
\multicolumn{1}{c}{\textbf{SSIM}$\uparrow$} &
\multicolumn{1}{c}{\textbf{SAM}$\downarrow$} &
\multicolumn{1}{c}{\scriptsize{\textbf{ERGAS}$\downarrow$}} &
\multicolumn{1}{c}{\textbf{MAE}$\downarrow$} &
\multicolumn{1}{c}{\textbf{MSE}$\downarrow$} &
\multicolumn{1}{c}{\textbf{PSNR}$\uparrow$} &
\multicolumn{1}{c}{\textbf{SSIM}$\uparrow$} &
\multicolumn{1}{c}{\textbf{SAM}$\downarrow$} &
\multicolumn{1}{c}{\scriptsize{\textbf{ERGAS}$\downarrow$}} &
\multicolumn{1}{c}{\textbf{MAE}$\downarrow$} &
\multicolumn{1}{c}{\textbf{MSE}$\downarrow$} &
\\ \midrule

\multirow{5}{*}[-1.5ex]{\rotatebox{90}{MFSR}}
& \multicolumn{2}{l}{\hspace{3em}MF-SRCNN} & 36.8263 & 0.8767 & 2.6776 & 9.3946 & 1.4681 & 9.6654 & 39.0772 & 0.8964 & 2.4051 & 5.5574 & 1.0371 & 4.0053 & 1778.77K \\
& \multicolumn{2}{l}{\hspace{3em}HighRes-Net} & 37.0815 & 0.8763 & 2.2503 & 9.1406 & 1.4498 & 9.5003 & 39.7523 & 0.9025 & 1.6448 & 5.1644 & 0.9519 & 3.4906 & 1627.98K \\
& \multicolumn{2}{l}{\hspace{3em}RAMS} & 37.1946 & 0.8793 & 2.3063 & 8.8203 & 1.4097 & 9.1261 & 40.3275 & 0.9101 & 1.4961 & 4.7717 & 0.8771 & 3.0101 & 338.06K \\
& \multicolumn{2}{l}{\hspace{3em}TR-MISR} & 37.0014 & 0.8778 & 2.3378 & 9.2222 & 1.4283 & 9.2299 & 39.5560 & 0.9005 & 1.7136 & 5.2308 & 0.9714 & 3.6965 & 470.35K \\ \cmidrule(lr){2-16}
& \multicolumn{2}{c}{\textbf{Average}} & 37.0260 & 0.8775 & 2.3930 & 9.1444 & 1.4390 & 9.3804 & 39.6783 & 0.9024 & 1.8149 & 5.1811 & 0.9594 & 3.5506 &  \\ \midrule

\multirow{6}{*}{\rotatebox{90}{Pansharpen}} 
& \multicolumn{2}{l}{\hspace{3em}PNN} & 46.4287 & 0.9877 & 2.1886 & 2.6345 & 0.4341 & 0.5246 & 47.5420 & 0.9886 & 1.9289 & 2.3656 & 0.4192 & 0.4289 & 76.04K \\
& \multicolumn{2}{l}{\hspace{3em}PanNet} & 45.6398 & 0.9843 & 2.3978 & 3.0284 & 0.4836 & 0.6414 & 48.0819 & 0.9900 & 1.8248 & 1.9513 & 0.3376 & 0.3108 & 308.68K \\
& \multicolumn{2}{l}{\hspace{3em}U2Net} & 46.8601 & 0.9859 & 2.2141 & 2.6720 & 0.4182 & 0.4750 & 47.4352 & 0.9910 & 1.9393 & 2.1106 & 0.3707 & 0.3519 & 632.81K \\
& \multicolumn{2}{l}{\hspace{3em}Pan-Mamba} & 45.7723 & 0.9861 & 2.4807 & 2.8916 & 0.4606 & 0.5840 & 47.9158 & 0.9882 & 1.7338 & 2.0379 & 0.3451 & 0.3324 & 479.48K \\
& \multicolumn{2}{l}{\hspace{3em}ARConv} & 46.6602 & 0.9864 & 2.2151 & 2.7638 & 0.4275 & 0.4944 & 46.8972 & 0.9873 & 2.0434 & 2.2976 & 0.3976 & 0.3897 & 15922.42K \\ \cmidrule(lr){2-16}
& \multicolumn{2}{c}{\textbf{Average}} & 46.2722 & 0.9861 & 2.2993 & 2.7981 & 0.4448 & 0.5439 & 47.5744 & 0.9890 & 1.8940 & 2.1526 & 0.3740 & 0.3627 &  \\ \midrule

\multirow{23}{*}[-6ex]{\rotatebox{90}{\textbf{\textit{SatFusion}}}} 
& \multicolumn{2}{l}{{\textit{MFIF}$_{fusion}$ \quad \textit{MSIF}$_{fusion}$}} & \multicolumn{12}{c}{} & \\ \cmidrule(lr){2-16}
& \multicolumn{1}{l}{\multirow{5}{*}{MF-SRCNN}} & PNN & 46.9912 & 0.9903 & 1.9501 & 2.4087 & 0.4037 & 0.4621 & 48.9561 & 0.9945 & 1.7296 & 1.8634 & 0.3113 & 0.2782 & 1853.20K \\
& & PanNet & 46.7910 & 0.9896 & 2.1471 & 2.5141 & 0.4100 & 0.4779 & 47.5788 & 0.9911 & 2.0159 & 2.2044 & 0.3735 & 0.3838 & 2085.84K \\
& & U2Net & 47.0066 & 0.9888 & 2.1659 & 2.5733 & 0.4130 & 0.4813 & 47.7126 & 0.9898 & 1.9826 & 2.1318 & 0.3710 & 0.3820 & 2409.97K \\
& & Pan-Mamba & 46.5703 & 0.9912 & 2.4536 & 2.4990 & 0.4190 & 0.5087 & 47.5974 & 0.9930 & 2.2166 & 2.1428 & 0.3647 & 0.3969 & 2256.63K \\
& & ARConv & 47.1350 & 0.9878 & 1.9386 & 2.5414 & 0.4059 & 0.4595 & 47.6784 & 0.9895 & 1.7688 & 2.1659 & 0.3764 & 0.3811 & 17699.58K \\ \cmidrule(lr){2-16}
& \multicolumn{1}{l}{\multirow{5}{*}{HighRes-Net}} & PNN & 46.9310 & 0.9887 & 2.0203 & 2.5345 & 0.4041 & 0.4840 & 49.5682 & 0.9922 & 1.6734 & 1.6948 & 0.2925 & 0.2494 & 1702.42K \\
& & PanNet & 47.0376 & 0.9890 & 2.0267 & 2.4888 & 0.3978 & 0.4517 & 49.7340 & 0.9947 & 1.6652 & 1.6614 & 0.2855 & 0.2419 & 1935.06K \\
& & U2Net & 47.3020 & 0.9895 & 1.8844 & 2.4569 & 0.3981 & 0.4469 & 49.1785 & 0.9932 & 1.6735 & 1.7584 & 0.3130 & 0.2799 & 2259.18K \\
& & Pan-Mamba & 46.5283 & 0.9906 & 2.3986 & 2.5027 & 0.4196 & 0.5064 & 48.5836 & 0.9952 & 1.9645 & 1.9224 & 0.3262 & 0.2973 & 2105.85K \\
& & ARConv & 47.1686 & 0.9890 & 1.9508 & 2.7859 & 0.4038 & 0.4567 & 48.2011 & 0.9903 & 1.7761 & 1.9991 & 0.3553 & 0.3469 & 17548.80K \\ \cmidrule(lr){2-16}
& \multicolumn{1}{l}{\multirow{5}{*}{RAMS}} & PNN & 47.1100 & 0.9907 & 1.9679 & 2.3450 & 0.3971 & 0.4610 & 50.0541 & 0.9928 & 1.4765 & 1.5354 & 0.2708 & 0.2126 & 412.49K \\
& & PanNet & 47.0404 & 0.9888 & 1.9705 & 2.4689 & 0.3993 & 0.4493 & 50.2041 & 0.9968 & 1.5331 & 1.5722 & 0.2647 & 0.2011 & 645.13K \\
& & U2Net & 47.5786 & 0.9913 & 1.8592 & 2.4918 & 0.3842 & 0.4174 & 48.6907 & 0.9902 & 1.7652 & 1.8493 & 0.3121 & 0.2754 & 969.26K \\
& & Pan-Mamba & 47.0081 & 0.9924 & 2.1994 & 2.5279 & 0.3987 & 0.4556 & 49.7599 & 0.9916 & 1.6422 & 1.6407 & 0.2783 & 0.2333 & 815.92K \\
& & ARConv & 47.1412 & 0.9877 & 1.9462 & 2.6387 & 0.4076 & 0.4584 & 48.7653 & 0.9892 & 1.6328 & 1.8581 & 0.3229 & 0.2757 & 16258.87K \\ \cmidrule(lr){2-16}
& \multicolumn{1}{l}{\multirow{5}{*}{TR-MISR}} & PNN & 46.7560 & 0.9896 & 2.1167 & 2.5260 & 0.4174 & 0.4983 & 49.5335 & 0.9929 & 1.6156 & 1.6512 & 0.2914 & 0.2439 & 544.78K \\
& & PanNet & 46.9719 & 0.9898 & 1.9917 & 2.5176 & 0.4031 & 0.4688 & 49.6781 & 0.9928 & 1.6198 & 1.6158 & 0.2875 & 0.2510 & 777.42K \\
& & U2Net & 47.6068 & 0.9890 & 1.9046 & 2.4587 & 0.3814 & 0.4150 & 48.4520 & 0.9903 & 1.9849 & 1.9062 & 0.3373 & 0.3309 & 1101.55K \\
& & Pan-Mamba & 46.7936 & 0.9884 & 2.0236 & 2.4335 & 0.4112 & 0.4923 & 49.5882 & 0.9965 & 1.7126 & 1.6903 & 0.2892 & 0.2577 & 948.22K \\
& & ARConv & 47.4052 & 0.9889 & 1.9852 & 2.5244 & 0.3914 & 0.4334 & 48.2372 & 0.9913 & 1.7387 & 1.9963 & 0.3505 & 0.3320 & 16391.17K \\ \cmidrule(lr){2-16}
& \multicolumn{2}{c}{\textbf{Average}} & \underline{47.0437} & \textbf{0.9896} & \underline{2.0451} & \underline{2.5119} & \underline{0.4033} & \underline{0.4642} & \underline{48.8876} & \underline{0.9924} & \underline{1.7594} & \underline{1.8430} & \underline{0.3187} & \underline{0.2926} &  \\ \midrule

\multirow{7}{*}[-2ex]{\rotatebox{90}{\textbf{\textit{SatFusion*}}}} 
& \multicolumn{2}{c}{{\textit{MSIF}$_{fusion}$}} & \multicolumn{12}{c}{} & \\ \cmidrule(lr){2-16}
& \multicolumn{2}{l}{\hspace{3.5em}PNN} & 47.2238 & 0.9898 & 1.9151 & 2.3275 & 0.3960 & 0.4472 & 49.8449 & 0.9935 & 1.6380 & 1.6256 & 0.2823 & 0.2414 & 545.48K \\
& \multicolumn{2}{l}{\hspace{3.5em}PanNet} & 47.3154 & 0.9875 & 1.9136 & 2.3372 & 0.3881 & 0.4359 & 48.9527 & 0.9926 & 1.7839 & 1.7906 & 0.3139 & 0.2933 & 778.12K \\
& \multicolumn{2}{l}{\hspace{3.5em}U2Net} & 47.7973 & 0.9878 & 1.8035 & 2.2640 & 0.3789 & 0.4147 & 49.4695 & 0.9911 & 1.6273 & 1.6910 & 0.2991 & 0.2626 & 1102.24K \\
& \multicolumn{2}{l}{\hspace{3.5em}Pan-Mamba} & 47.0228 & 0.9872 & 2.0780 & 2.3900 & 0.3986 & 0.4586 & 49.7388 & 0.9960 & 1.7369 & 1.6620 & 0.2829 & 0.2408 & 948.91K \\
& \multicolumn{2}{l}{\hspace{3.5em}ARConv} & 47.5509 & 0.9879 & 1.8447 & 2.3398 & 0.3840 & 0.4159 & 48.7734 & 0.9908 & 1.7211 & 1.8228 & 0.3289 & 0.3062 & 16391.85K \\ \cmidrule(lr){2-16}
& \multicolumn{2}{c}{\textbf{Average}} & \textbf{47.3820} & \underline{0.9880} & \textbf{1.9110} & \textbf{2.3317} & \textbf{0.3891} & \textbf{0.4345} & \textbf{49.3559} & \textbf{0.9928} & \textbf{1.7014} & \textbf{1.7184} & \textbf{0.3014} & \textbf{0.2689} &  \\ 
\bottomrule
\end{tabular}
}
\par\smallskip
\raggedright \small \hspace{2em} \textbf{Bold} / \underline{Underline}: Best/second best among group averages. 
\end{table*}

\section{Details of Experimental Parameter Settings}
\label{app:settings}

To guarantee fair and reproducible comparisons, our \textit{SatFusion} variants and all baseline methods are evaluated under strictly consistent data and training configurations. This section details the specific hyperparameter settings employed across our experiments. All evaluations are executed on a server equipped with eight NVIDIA RTX 4090 GPUs.

\textbf{Configurations on the WorldStrat Dataset:}
Following the official WorldStrat benchmark~\cite{cornebise2022open}, we set the spatial dimensions to $\gamma H=\gamma W=156$ for the HR targets and $H=W=50$ for the LR inputs, corresponding to an effective spatial upscaling factor of $\gamma \approx 3$. The input MS frames contain $C=3$ spectral channels, and the sequence length is fixed to $k=8$. In our feature extraction and fusion modules, the internal channel capacities are set to $C_{enc} = 128$ and $C_{fus} = 128$. The internal sub-pixel convolution block within $MFIF_{decode}$ utilizes an upscaling factor of $\gamma' = 2$, followed by the exact interpolation-based resizing step defined in Appendix~\ref{app:architecture} to ensure strict spatial alignment with the PAN image. During training, we utilize the Adam optimizer paired with a Cosine Annealing Warm Restarts scheduler~\cite{loshchilov2016sgdr}. The batch size is set to $8$, and the models are trained for a maximum of 20 epochs.

\textbf{Configurations on Simulated Datasets (WV3, QB, and GF2):}
For experiments on the simulated satellite datasets, we adopt the standard configurations provided by the DLPan-Toolbox~\cite{deng2022machine}. Taking the WV3 dataset as a representative example, the training and validation patches are cropped to spatial dimensions of $H = W = 16$ for the LR inputs and $\gamma H=\gamma W=64$ for the HR targets (yielding $\gamma=4$). The MS imagery contains $C=8$ spectral channels, and the multi-frame sequence length is configured as $k=8$. During the testing phase, the spatial dimensions are expanded to $H = W = 64$ and $\gamma H=\gamma W = 256$. The internal channel capacities remain fixed at $C_{enc} = C_{fus} = 128$.
To generate the realistically degraded multi-frame sequences via our physics-inspired pipeline (Algorithm~\ref{alg:synthesis}), we apply a consistent set of degradation parameters across these datasets. Specifically, the sub-pixel spatial shift range is set to $\Delta \in [-1, 1]$. The standard deviation for the Gaussian blur, which simulates the sensor PSF/MTF effects, is uniformly sampled from $\sigma \in [1.2, 1.8]$. To emulate realistic photon capture noise, the Poisson shot noise gain is fixed at $g = 5 \times 10^3$, and the Gaussian readout noise standard deviation is set to $\sigma_r = 0.001$.

While the data dimensions are uniform across models, the specific training hyperparameters (e.g., total epochs, batch size, and optimizer) vary depending on the instantiated Pansharpening components to match their original optimal settings. Table~\ref{tab:parameters} summarizes the precise training configurations for the representative baselines. To instantiate \textit{SatFusion}, we integrate MFSR components into the $MFIF_{fusion}$ interface. For a fair comparison, our models strictly inherit the original training hyperparameters (e.g., optimizers, epochs) of their corresponding Pansharpening baselines from the DLPan-Toolbox~\cite{deng2022machine}, modifying only the network architecture and joint loss formulation.

\begin{table*}[htbp]
\centering
\caption{Exhaustive experimental metrics on WV3, GF2, and QB simulated datasets.}
\label{tab:full_results_wv3}
\small
\renewcommand{\arraystretch}{0.75}
\setlength{\tabcolsep}{4pt}

\resizebox{\textwidth}{!}{
\begin{tabular}{lll cccc cccc cccc}
\toprule
\multicolumn{3}{c}{\multirow{2}{*}{\textbf{Methods}}} & \multicolumn{4}{c}{\textbf{WV3}} & \multicolumn{4}{c}{\textbf{GF2}} & \multicolumn{4}{c}{\textbf{QB}} \\ 
\cmidrule(lr){4-7} \cmidrule(lr){8-11} \cmidrule(lr){12-15}
\multicolumn{3}{c}{} & \textbf{PSNR}$\uparrow$ & \textbf{SSIM}$\uparrow$ & \textbf{SAM}$\downarrow$ & \scriptsize{\textbf{ERGAS}$\downarrow$} &
\textbf{PSNR}$\uparrow$ & \textbf{SSIM}$\uparrow$ & \textbf{SAM}$\downarrow$ & \scriptsize{\textbf{ERGAS}$\downarrow$} &
\textbf{PSNR}$\uparrow$ & \textbf{SSIM}$\uparrow$ & \textbf{SAM}$\downarrow$ & \scriptsize{\textbf{ERGAS}$\downarrow$} \\ 
\midrule

\multirow{7}{*}[-0.7ex]{\rotatebox{90}{Pansharpen}} 
& \multicolumn{2}{l}{\hspace{3.5em}PNN} & 36.5340 & 0.9548 & 4.2758 & 3.6505 & 41.9402 & 0.9696 & 1.5319 & 1.4494 & 36.0032 & 0.9264 & 5.1423 & 5.7539 \\
& \multicolumn{2}{l}{\hspace{3.5em}DiCNN} & 37.1690 & 0.9611 & 4.0397 & 3.4236 & 42.4487 & 0.9729 & 1.4386 & 1.3750 & 36.2339 & 0.9305 & 4.9892 & 5.6132 \\
& \multicolumn{2}{l}{\hspace{3.5em}MSDCNN} & 35.9721 & 0.9454 & 4.7528 & 3.9338 & 42.0847 & 0.9702 & 1.5241 & 1.4278 & 35.8757 & 0.9254 & 5.1286 & 5.8496 \\
& \multicolumn{2}{l}{\hspace{3.5em}DRPNN} & 37.1089 & 0.9603 & 4.1000 & 3.4253 & 43.1093 & 0.9760 & 1.3330 & 1.2747 & 37.3074 & 0.9436 & 4.7667 & 4.9032 \\
& \multicolumn{2}{l}{\hspace{3.5em}FusionNet} & 37.5678 & 0.9634 & 3.8872 & 3.2372 & 42.7230 & 0.9740 & 1.3562 & 1.3319 & 36.8057 & 0.9379 & 4.9236 & 5.1991 \\
& \multicolumn{2}{l}{\hspace{3.5em}U2Net} & 38.0416 & 0.9678 & 3.6772 & 3.0081 & 43.1198 & 0.9763 & 1.2491 & 1.2930 & 37.7626 & 0.9479 & 4.6238 & 4.6672 \\
\cmidrule(lr){2-15}
& \multicolumn{2}{c}{\textbf{Average}} & 37.0656 & 0.9588 & 4.1221 & 3.4464 & 42.5710 & 0.9732 & 1.4055 & 1.3586 & 36.6648 & 0.9353 & 4.9290 & 5.3310 \\
\midrule

\multirow{26}{*}[-8ex]{\rotatebox{90}{\textbf{\textit{SatFusion}}}}  & \multicolumn{2}{l}{$MSIF_{fusion} \quad MFIF_{fusion}$} & & & & & & & & & & & & \\ \cmidrule(lr){2-15}
& \multirow{4}{*}{PNN} & MF-SRCNN & 36.8196 & 0.9608 & 4.3168 & 3.5191 & 41.8845 & 0.9724 & 1.6167 & 1.4753 & 36.0371 & 0.9279 & 5.1139 & 5.7658 \\
& & HighRes-Net & 37.0614 & 0.9617 & 4.1264 & 3.4345 & 42.3180 & 0.9731 & 1.5600 & 1.4153 & 36.0488 & 0.9282 & 5.1232 & 5.7598 \\
& & RAMS  & 37.1502 & 0.9607 & 4.0999 & 3.4178 & 42.7533 & 0.9736 & 1.4507 & 1.3559 & 36.5932 & 0.9331 & 4.9491 & 5.3915 \\
& & TR-MISR & 36.8628 & 0.9619 & 4.1233 & 3.4616 & 42.6706 & 0.9736 & 1.4440 & 1.3639 & 36.1081 & 0.9300 & 5.0542 & 5.7149 \\ 
\cmidrule(lr){2-15}

& \multirow{4}{*}{DiCNN} & MF-SRCNN & 37.8822 & 0.9682 & 3.6145 & 3.1235 & 42.7883 & 0.9761 & 1.2998 & 1.3574 & 37.5458 & 0.9486 & 4.4724 & 4.7414 \\
& & HighRes-Net & 38.6588 & 0.9714 & 3.3471 & 2.8626 & 43.4098 & 0.9772 & 1.1542 & 1.2752 & 37.8274 & 0.9504 & 4.3736 & 4.6255 \\
& & RAMS & 38.4515 & 0.9716 & 3.3458 & 2.9224 & 43.2552 & 0.9764 & 1.1565 & 1.2817 & 38.3380 & 0.9555 & 4.2156 & 4.2813 \\
& & TR-MISR & 38.3025 & 0.9709 & 3.4684 & 2.9962 & 44.3630 & 0.9806 & 1.0832 & 1.1274 & 37.5241 & 0.9484 & 4.3801 & 4.7971 \\ 
\cmidrule(lr){2-15}

& \multirow{4}{*}{MSDCNN} & MF-SRCNN & 36.6887 & 0.9590 & 4.2955 & 3.5976 & 42.4895 & 0.9743 & 1.4321 & 1.3893 & 36.1084 & 0.9335 & 4.7937 & 5.6581 \\
& & HighRes-Net & 36.7270 & 0.9594 & 4.3319 & 3.5825 & 43.0057 & 0.9760 & 1.3138 & 1.3177 & 35.9599 & 0.9324 & 4.8643 & 5.7786 \\
& & RAMS & 36.8621 & 0.9605 & 4.1827 & 3.5228 & 43.1134 & 0.9760 & 1.3203 & 1.3107 & 36.2425 & 0.9341 & 4.7583 & 5.5703 \\
& & TR-MISR & 36.8286 & 0.9605 & 4.1982 & 3.5319 & 42.9072 & 0.9751 & 1.3560 & 1.3379 & 36.3045 & 0.9349 & 4.7946 & 5.5322 \\ 
\cmidrule(lr){2-15}

& \multirow{4}{*}{DRPNN} & MF-SRCNN & 37.0290 & 0.9632 & 4.0149 & 3.4516 & 43.6850 & 0.9789 & 1.2164 & 1.2196 & 37.7417 & 0.9513 & 4.4960 & 4.6487 \\
& & HighRes-Net & 37.4433 & 0.9654 & 3.7812 & 3.3273 & 44.3259 & 0.9807 & 1.1437 & 1.1412 & 38.1620 & 0.9551 & 4.3119 & 4.3766 \\
& & RAMS & 37.5732 & 0.9655 & 3.7671 & 3.2905 & 44.6619 & 0.9819 & 1.1479 & 1.0826 & 38.2824 & 0.9552 & 4.2797 & 4.3140 \\
& & TR-MISR & 37.4104 & 0.9650 & 3.8038 & 3.3520 & 45.0867 & 0.9829 & 1.0532 & 1.0317 & 38.2458 & 0.9559 & 4.2735 & 4.3351 \\ 
\cmidrule(lr){2-15}

& \multirow{4}{*}{FusionNet} & MF-SRCNN & 37.8537 & 0.9688 & 3.7209 & 3.1390 & 43.3958 & 0.9776 & 1.1749 & 1.2756 & 37.9621 & 0.9538 & 4.3859 & 4.4830 \\
& & HighRes-Net & 38.5889 & 0.9728 & 3.3880 & 2.8749 & 43.7922 & 0.9782 & 1.1204 & 1.2290 & 38.3800 & 0.9569 & 4.1744 & 4.2750 \\
& & RAMS & 38.3149 & 0.9708 & 3.3861 & 2.9825 & 43.6333 & 0.9766 & 1.1063 & 1.2347 & 38.4529 & 0.9574 & 4.1426 & 4.2295 \\
& & TR-MISR & 38.6914 & 0.9729 & 3.3040 & 2.8482 & 44.5058 & 0.9816 & 1.1027 & 1.1054 & 38.4834 & 0.9581 & 4.1139 & 4.2345 \\
\cmidrule(lr){2-15}

& \multirow{4}{*}{U2Net} & MF-SRCNN & 38.1168 & 0.9698 & 3.5777 & 2.9076 & 43.1605 & 0.9783 & 1.2364 & 1.2449 & 37.8251 & 0.9490 & 4.6643 & 4.7903 \\
& & HighRes-Net & 37.3474 & 0.9641 & 4.0147 & 3.3194 & 43.7863 & 0.9793 & 1.1252 & 1.2076 & 37.9579 & 0.9517 & 4.5857 & 4.5578 \\
& & RAMS & 39.2459 & 0.9641 & 4.0147 & 3.3194 & 43.7863 & 0.9793 & 1.1252 & 1.2076 & 37.9579 & 0.9517 & 4.5857 & 4.5578 \\
& & TR-MISR & 39.1302 & 0.9753 & 3.0939 & 2.6848 & 44.2919 & 0.9815 & 1.0925 & 1.1344 & 38.6255 & 0.9590 & 4.1474 & 4.1469\\ 
\cmidrule(lr){2-15}

& \multicolumn{2}{c}{\textbf{Average}} & \underline{37.7100} & \underline{0.9665} & \underline{3.7672} & \underline{3.2003} & \underline{43.5206} & \underline{0.9778} & \underline{1.2373} & \underline{1.2475} & \underline{37.4679} & \underline{0.9466} & \underline{4.5274} & \underline{4.8447} \\  
\midrule

\multirow{7}{*}[-3ex]{\rotatebox{90}{\textbf{\textit{SatFusion*}}}} & \multicolumn{2}{c}{$MSIF_{fusion}$} & & & & & & & & & & & & \\  \cmidrule(lr){2-15}
& \multicolumn{2}{l}{\hspace{3.5em}PNN} & 37.3092 & 0.9627 & 4.0141 & 3.3731 & 42.8426 & 0.9744 & 1.4201 & 1.3407 & 36.2618 & 0.9307 & 5.0586 & 5.6360 \\
& \multicolumn{2}{l}{\hspace{3.5em}DiCNN} & 38.3609 & 0.9710 & 3.4074 & 2.9532 & 45.8234 & 0.9864 & 0.9959 & 0.9337 & 38.6826 & 0.9595 & 4.1140 & 4.1168 \\
& \multicolumn{2}{l}{\hspace{3.5em}MSDCNN} & 36.8714 & 0.9584 & 4.4364 & 3.5939 & 43.0315 & 0.9761 & 1.3360 & 1.3027 & 36.4599 & 0.9350 & 4.7822 & 5.5434 \\
& \multicolumn{2}{l}{\hspace{3.5em}DRPNN} & 37.4506 & 0.9655 & 3.8028 & 3.3206 & 44.9421 & 0.9826 & 1.0710 & 1.0491 & 38.1898 & 0.9553 & 4.3033 & 4.3701 \\
& \multicolumn{2}{l}{\hspace{3.5em}FusionNet} & 39.0767 & 0.9748 & 3.1360 & 2.7155 & 45.5417 & 0.9862 & 1.0247 & 0.9738 & 38.7960 & 0.9605 & 4.0505 & 4.0650 \\
& \multicolumn{2}{l}{\hspace{3.5em}U2Net} & 39.1022 & 0.9762 & 3.1035 & 2.7015 & 45.3869 & 0.9847 & 0.9989 & 1.0013 & 38.2583 & 0.9572 & 4.3039 & 4.2850 \\
\cmidrule(lr){2-15}
& \multicolumn{2}{c}{\textbf{Average}} & \textbf{38.0285} & \textbf{0.9681} & \textbf{3.6500} & \textbf{3.1096} & \textbf{44.5947} & \textbf{0.9817} & \textbf{1.1411} & \textbf{1.1002} & \textbf{37.7747} & \textbf{0.9497} & \textbf{4.4346} & \textbf{4.6694}  \\
\bottomrule
\end{tabular}
}
\par\smallskip
\raggedright \small \hspace{2em} \textbf{Bold} / \underline{Underline}: Best/second best among group averages. 
\end{table*}

\section{Exhaustive Quantitative Results for WorldStrat Modular Combinations}
\label{app:full_results}
As discussed in Section 4.3.1 of the main text, our proposed unified framework allows seamless integration of various multi-frame feature aggregation strategies ($MFIF_{fusion}$) and multi-source fusion mechanisms ($MSIF_{fusion}$). 

Table~\ref{tab:full_results} provides the exhaustive quantitative evaluation results across all combinations of these modules on both the real-world and realistically simulated WorldStrat datasets. The exhaustive testing includes 20 architectural variants for \textit{SatFusion} (combining 4 MFSR operators and 5 Pansharpening operators) and 5 architectural variants for \textit{SatFusion*} (combining our proposed PAN-guided Transformer with 5 Pansharpening operators). In addition, we report the parameter count (\#Params) for each specific instantiation. These comprehensive results demonstrate that our framework consistently yields performance improvements regardless of the specific underlying modular choice, confirming its robustness and high extensibility.

\section{Exhaustive Quantitative Results for WV3, GF2, and QB Modular Combinations}
\label{app:full_results_wv3}
Table~\ref{tab:full_results_wv3} details the performance of all investigated modular combinations of \textit{SatFusion} and \textit{SatFusion*} on the WV3, GF2, and QB simulated datasets, supplementing the summarized performance presented in Table~\ref{tab:2} of the main manuscript.

\begin{figure*}[htbp]
    \centering
    \includegraphics[width=0.91\linewidth]{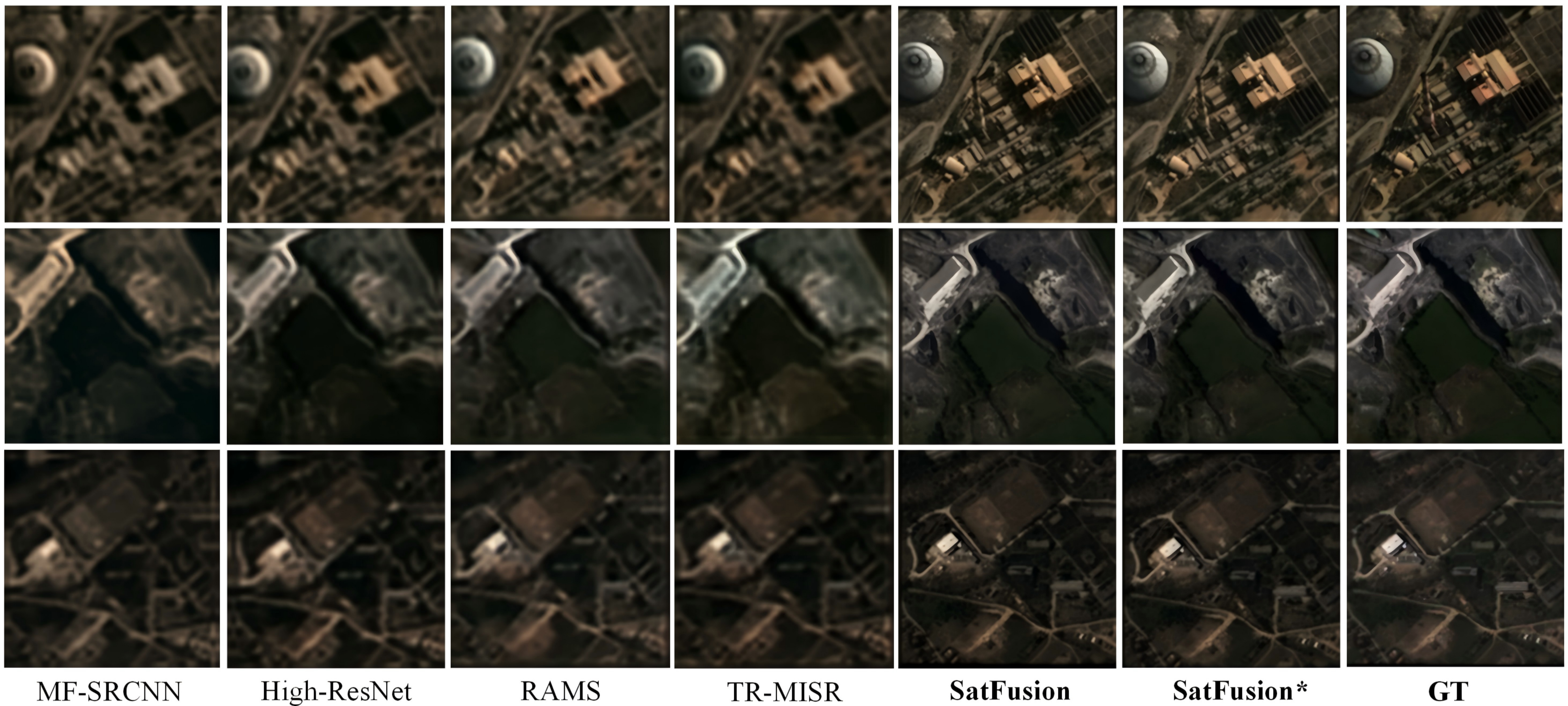}
    \caption{
        Qualitative comparison of different fusion methods on the WorldStrat dataset. By effectively integrating fine-grained spatial details from the PAN image, \textit{SatFusion} and \textit{SatFusion*} produce visually superior reconstructions with sharper structures and clearer textures compared to MFSR methods that rely solely on multi-frame information.
    }
    \label{fig:6_app}
\end{figure*}
\begin{figure*}[htbp]
    \centering
    \includegraphics[width=0.93\linewidth]{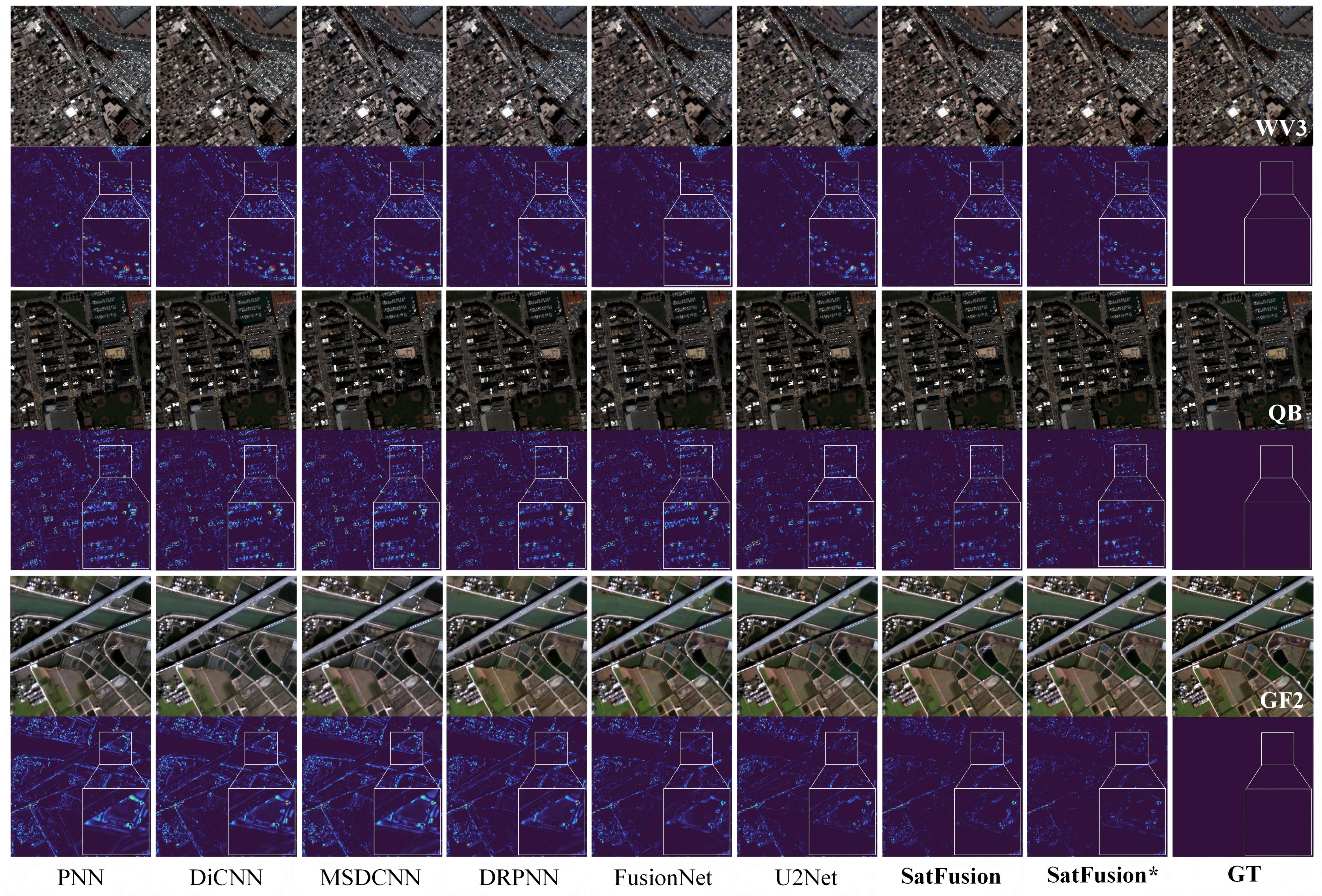}
    \caption{
        Qualitative comparison on the simulated data (WV3, QB, and GF2). The error maps visually demonstrate that \textit{SatFusion*} yields the lowest reconstruction discrepancy with respect to the GT. Benefiting from the complementary information across multiple frames, our method successfully suppresses input artifacts while accurately recovering spatial details.
    }
    \label{fig:7_app}
\end{figure*}

\section{Qualitative Results on WorldStrat}
\label{app:qual_worldstrat}
To complement the quantitative results presented in Section 4.3.1 of the main manuscript, we provide visual comparisons of the reconstructed images on the WorldStrat dataset. 
As illustrated in Fig.~\ref{fig:6_app}, MFSR methods generally produce overly smooth outputs due to the absence of high-frequency structural guidance. In contrast, by effectively integrating fine-grained spatial details from the PAN image, both \textit{SatFusion} and \textit{SatFusion*} produce visually superior reconstructions. Our proposed unified methods explicitly exhibit sharper edge structures and more accurately restored local textures, corroborating the significant numerical improvements reported in the main text.

\section{Qualitative Results on WV3, QB, and GF2}
\label{app:qual_simulated}
To further demonstrate the robustness of our framework against real-world perturbations (e.g., sub-pixel misalignment and noise), we present qualitative comparisons on the physics-inspired simulated data (including WV3, QB, and GF2 datasets) complementing Section 4.3.2.
Fig.~\ref{fig:7_app} visualizes the fused images alongside their corresponding error maps with respect to the Ground Truth (GT). Our method effectively leverages complementary information across multiple frames to enhance fusion quality. Benefiting from the multi-frame modeling guided by PAN structural priors, \textit{SatFusion*} delivers reconstructions with lower error magnitudes and superior perceptual clarity.

\section{Real-World Implications}
\label{app:implication}
Beyond the quantitative and qualitative improvements demonstrated in the main manuscript, our unified framework offers significant practical advantages for real-world Satellite Internet of Things (Sat-IoT) deployments.

\textbf{Reliable High-Fidelity Perception.} 
In practical Earth observation, hardware sensor limitations often exacerbate the conflict between the acquisition of low-quality redundant data and the downstream demand for high-fidelity imagery. By synergistically integrating multi-frame temporal complementarity and multi-source spatial priors, our framework bridges this gap. Unlike traditional Pansharpening methods that rely on fragile single-image interpolation, our method achieves alignment implicitly within a deep high-resolution feature space. This ensures highly stable and reliable reconstructions even under severe satellite jitter, sensor noise, or atmospheric interference, providing trustworthy inputs for downstream analytical tasks.

\begin{figure}[htbp]
    \centering
    \includegraphics[width=0.8\linewidth]{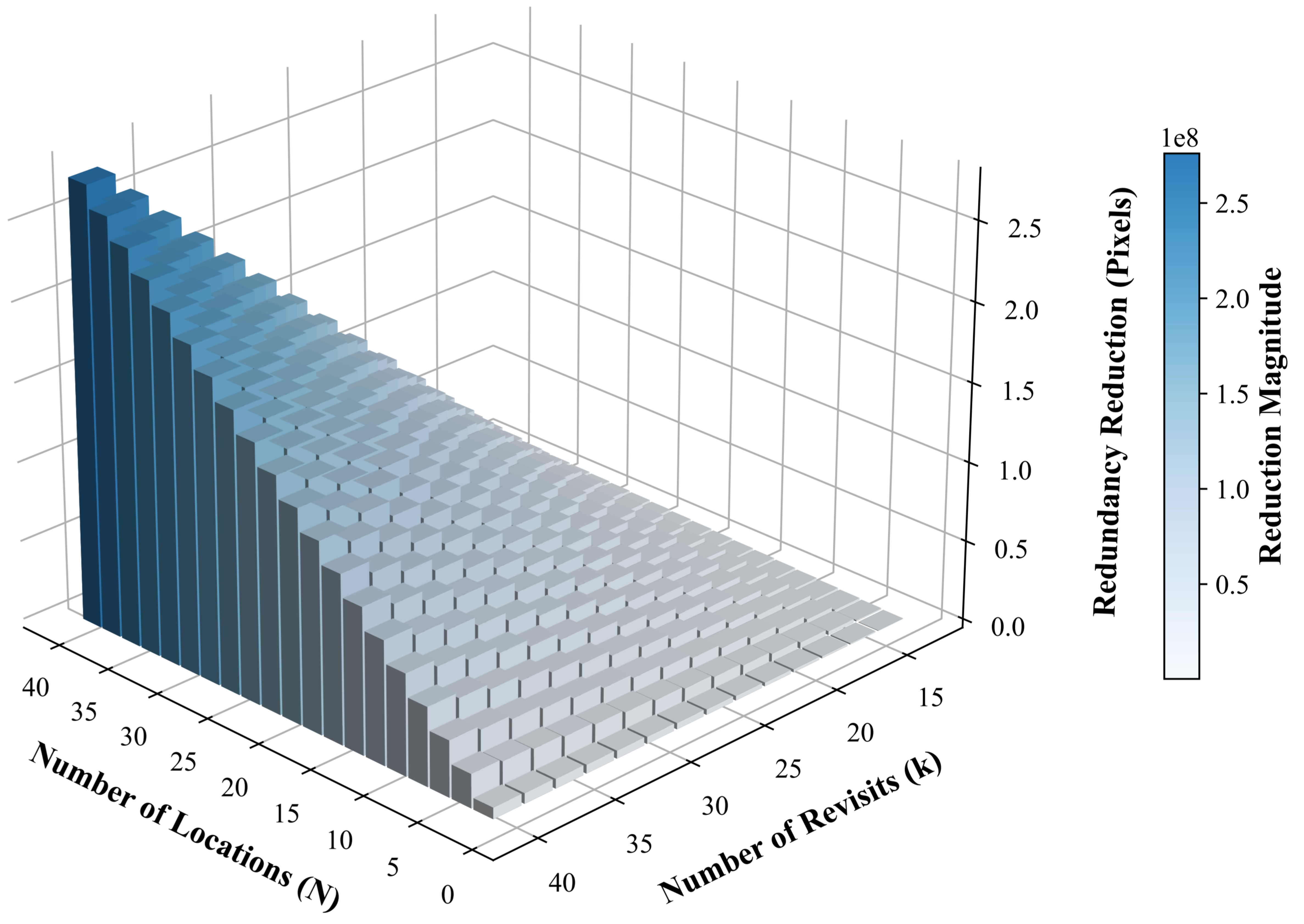}
    \caption{
        Difference in data volume between input and output images ($\Delta D = D_{\text{in}} - D_{\text{out}}$), under the setting $H=W=256$, $C=4$, and $\gamma=4$. The positive compression benefit becomes increasingly pronounced as the number of available frames $k$ grows in dense Sat-IoT scenarios.
    }
    \label{fig:12}
\end{figure}

\textbf{Bandwidth and Storage Efficiency.} 
In Sat-IoT networks, managing and transmitting raw, highly overlapping temporal sequences imposes a massive burden on system bandwidth and storage capacities. Our approach consolidates multiple low-quality frames into a single high-quality representation, naturally yielding data compression benefits. Assuming each pixel per channel occupies one unit of storage space, and letting $N$ denote the spatial footprint, the raw input data volume comprising $k$ LR MS frames and one HR PAN image is:
\begin{equation}
D_{\text{in}} = N \times (k \cdot H \cdot W \cdot C + \gamma H \cdot \gamma W \cdot 1).
\end{equation}
The output data volume of the single fused HR MS image is:
\begin{equation}
D_{\text{out}} = N \times (\gamma H \cdot \gamma W \cdot C).
\end{equation}
Because the fused output possesses enhanced spatial resolution ($\gamma$) and spectral depth ($C$), $D_{\text{out}}$ can initially exceed $D_{\text{in}}$ for small $k$. However, in dense revisit scenarios typical of modern Sat-IoT constellations, $k$ is usually substantial. As illustrated in Fig.~\ref{fig:12}, the system transitions into a net compression regime ($D_{\text{in}} - D_{\text{out}} > 0$) once $k$ surpasses a specific threshold. This advantage becomes increasingly prominent as $k$ grows, allowing the overall system to significantly reduce data payloads and archiving costs while simultaneously delivering superior image fidelity.

\end{document}